\documentclass[fleqn,usenatbib]{mnras}
\usepackage{newtxtext,newtxmath}

\usepackage[T1]{fontenc}

\newcommand{\kms}{\,km\,s$^{-1}$} 
\newcommand{\halph}{H$\alpha$ }
\newcommand{\halphend}{H$\alpha$}
\newcommand{\app}{\(\sim\)}
\newcommand{\he}{\ion{He}{i} $\lambda$6678 }
\newcommand{\hel}{\ion{He}{i} $\lambda$5875 }
\newcommand{\heend}{\ion{He}{i} $\lambda$6678}
\newcommand{\hei}{\ion{He}{i} $\lambda$4922 }
\newcommand{\heii}{\ion{He}{ii} $\lambda$5411 }
\newcommand{\hii}{\ion{H}{ii}\ }

\DeclareRobustCommand{\VAN}[3]{#2}
\let\VANthebibliography\thebibliography
\def\thebibliography{\DeclareRobustCommand{\VAN}[3]{##3}\VANthebibliography}

\usepackage{graphicx}	
\usepackage{amsmath}	

\title[The Oe Star VES 735]{Long-Term Monitoring of the Oe Star VES 735: Ope! Not So Quiet After All}

\author[Marshall \& Kerton]{
Brandon Marshall,$^{1}$\thanks{E-mail: marshallb@unk.edu}
and C. R. Kerton$^{2}$
\\
$^{1}$Department of Physics \& Astronomy, University of Nebraska at Kearney, 2504 9th Ave., Kearney, NE 68849, USA\\
$^{2}$Department of Physics \& Astronomy, Iowa State University, 2323 Osborn Dr., Ames IA 50011, USA
}

\date{Accepted 2024 January 17. Received 2024 January 15; in original form 2023 November 15}

\pubyear{2023}

\begin{document}
\label{firstpage}
\pagerange{\pageref{firstpage}--\pageref{lastpage}}
\maketitle

\begin{abstract}
Only 3--4 per cent of galactic O stars are observed to display the emission features representative of the OBe phenomenon, compared to galactic B stars, which display these characteristics in 25--35 per cent of B0 and B1 stars. We present new observations of the high-mass O star, VES 735, which confirms its classification as one of these rare emission-line stars. These are its first recorded observations that display strong spectroscopic variations in nearly 30 years of monitoring, with the \halph profile exhibiting a tenfold increase in emission compared to observations taken between 1996 and 2014 and having variations which show episodes of inflowing and outflowing material. These observations, coupled with photometric variations in the visible and infrared, show behavior that is consistent with the mass reservoir effect for viscous decretion discs. We propose that in 2015 VES 735 began an approximately three year event in which mass was being injected into the circumstellar environment followed by re-accretion towards the star. We also find evidence that the re-accretion may have been interrupted with another, smaller, mass-injection event based on observations in 2022 and 2023. Observational cadences ranging from hours to months show no evidence that VES 735 is part of a binary system, making it an ideal candidate for future observations to further investigate the evolution of high-mass stars and the OBe phenomenon as it pertains to their circumstellar environment.
\end{abstract}




\section{Introduction} \label{sec:intro}
Oe stars were first proposed as higher-mass analogs of Be stars by \citet{cl74}. Be stars are the result of the combination of a rapidly rotating B star and some mechanism (the `Be phenomenon'), possibly related to non-radial pulsations and/or magnetic fields, that results in material being transferred from the star into a circumstellar disc \citep{Riv13}. The, often double-peaked, Balmer line emission, which defines this class of stars, originates from the hot disc material. The spectra of Be stars can show three types of, not mutually exclusive, variability \citep{McL61,Gold16}: changes in the overall intensity of the emission lines, changes in the intensity of the `violet' and `red' components of the double-peaked emission (or V/R variation), and the appearance or disappearance of a `shell' (sharp and deep) absorption core. Oe stars are rare; for Galactic O stars of luminosity class of III--V, the Oe/O ratio has been found to be only a few per cent (\citet{neg2004}: 4 per cent, \citet{Gold16}: 3 per cent), whereas the ratio of Be/B stars was found to be 27 per cent for spectral type of B0 and 34 per cent for B1 \citep{Zorec97}.

VES~735 was initially classified as an O8.5 Ve star by \citet{KBM99} as part of a study of the surrounding \ion{H}{II} region KR~140, which is powered solely by VES~735. Spectroscopic monitoring of the star prior to 2019 had shown long-lived ($\gtrsim$ 20 years), double-peaked \halph emission of moderate intensity with slight variability across monthly and yearly timescales. This is in striking contrast to the behaviour of the more well-known late-type Oe star, $\zeta$ Oph, which exhibits more sporadic emission episodes \citep{niemela}. Given the steady nature of the observed emission, and the association with the fairly young \ion{H}{II} region KR~140, the disc around VES~735 was previously interpreted as a possible remnant accretion disk \citep{MK18,KBM99}. More recent observations of VES~735 display two of the spectroscopic variations characteristic of classical Oe and Be stars. The equivalent width of the \halph line has increased by a factor of nearly 25 since 2014, and the line has shown significant V/R variation on a monthly time scale. In addition, the double-peaked structure of the \halph line is now visible in multiple \ion{He}{I} lines, and large V/R variations are also visible in \heend. Given the presence of emission in \ion{He}{I}, in addition to the \halph emission, VES~735 is better classified as an Ope star \citep{Sota11}.

This paper presents an analysis of almost 30 years of spectroscopic and photometric observations of VES~735 with a particular emphasis on observations obtained over the past decade that trace the injection of material into a circumstellar disc and its subsequent evolution. Spectra taken with long (annual to decadal) and short-term (hourly to monthly) cadences do not show any evidence that VES~735 is a part of a binary system. This, along with its relatively young age (\app 2.5 Myr; \citealt{KBM99}) and proximity (\app 1.45 kpc; see Section 3.1), make it an ideal laboratory for studying high-mass stellar evolution and the OBe phenomenon. In Section \ref{sec:obs} we provide a summary of the spectroscopic and photometric observations and archival data used in our analysis. Section \ref{sec:spectype} reviews the evidence supporting the O-star classification for VES~735. Section \ref{sec:results} presents a detailed description of the available \halph and photometric measurements, and we discuss what the observations can tell us about VES~735 and the flow of material into and from its circumstellar disc. In Section \ref{sec:discuss} we discuss the possibility of binarity as well as stellar and disc parameters that can estimated from our observations. Conclusions are presented in Section \ref{sec:conclude}.

\section{Observations} 
\label{sec:obs}

To more clearly discuss the changes seen over time in the observational data we have defined four epochs based on the observed spectroscopic and photometric variation of VES~735 in Table~\ref{tab:Epochs}. These will be described in detail in Section~\ref{sec:results}, and the epoch nomenclature will be used for the remainder of the paper. 

\halph spectroscopy of VES~735 for this study was obtained using long-slit, Cassegrain-mounted spectrometers on the 1.9-m telescope at the David Dunlap Observatory (DDO), the 1.8-m Plaskett Telescope at the Dominion Astrophysical Observatory (DAO), the 2.3-m Wyoming Infrared Observatory (WIRO) telescope, and the 3.5-m telescope at the Apache Point Observatory (APO). The DDO and DAO spectra have spectral resolution of \app 0.5 and \app 0.7~\r{A} respectively, and the WIRO and APO spectra both have slightly lower spectral resolution of \app 2~\r{A}. We also obtained a higher resolution (\app 0.2~\r{A}) echelle spectrum using the 3.5-m APO telescope (only MJD 59728), and a lower resolution (\app 12~\r{A}) $K$-band spectrum using the Gemini Near-Infrared Spectrometer (GNIRS) on the 8-m Gemini North telescope. All spectroscopic data were reduced using {\sc iraf} reduction procedures from \citet{Massey} except the Gemini spectra, which were reduced using the Gemini/IRAF package (Ureka version 1.5.1) and GNIRS specific tools \citep{GNIRS}.

\begin{table}
    \caption{Epoch date ranges in UTC and MJD.}
    \centering
    \begin{tabular}{c|c|c}
    \hline
       Epoch  & Date (UTC) &  MJD \\
       \hline
       \hline
       I & 1996 September 24 -- 2014 September 1 & 50350 -- 56901 \\
       II & 2014 September 1 -- 2019 April 27 & 56901 -- 58600\\
       III & 2019 April 27 -- 2021 December 13 & 58600 -- 59570\\
       IV & 2021 December 13 -- 2023 January 11 & 59570 -- 59955\\
       \hline
    \end{tabular}
    \label{tab:Epochs}
\end{table}

A listing of all 
spectra obtained is provided in Table~\ref{tbl:obs}. The first four columns give the observatory code, the epoch in which the observation took place, the UTC start date/time of each observation, and the corresponding Modified Julian Date (MJD). In the case of the later DAO observations multiple spectra were obtained at 30 minute intervals during each observing session. As we observed no significant variation in the spectra at these time scales we created average spectra (listed in bold text). The remaining columns list properties of the double-peaked \halph line measured in each spectrum: the observed equivalent width ($EW$ in \r{A}), with negative indicating emission, the (non-continuum-subtracted) ratio of the height of the blue (`violet')  peak of the line to the red peak of the line ($V/R$), the velocity difference between the blue and red peak ($\Delta RV$ in km~s$^{-1}$), the maximum emission level relative to the continuum ($E/C$), and the full-width of the line at the continuum level ($W$ in \r{A}). Note that the equivalent widths reported have not been corrected for any underlying absorption of \halph (or Br$\gamma$ in the case of the GNIRS spectra). 

We also observed VES~735 with the FORCAST mid-infrared camera \citep{herter} on the the 2.5-m telescope aboard the Stratospheric Observatory for Infrared Astronomy \citep[SOFIA;][]{Young} as part of the Cycle 9 observing schedule (PlanID: $09\_0052$; see Table~\ref{tbl:obs} for observation times). FORCAST is a dual-channel camera with two $256 \times 256$ pixel focal plane arrays, each having a plate scale of 0.768 arcsec pixel\textsuperscript{-1}, operating at wavelengths from 2 -- 40~$\micron$. We obtained images at 7.7, 11.2, 25.3, and 34.8 $\micron$. The observations were processed by the \textit{SOFIA} Data Processing Team, and downloaded from the Infrared Science Archive (IRSA). Flux densities from \textit{SOFIA} were calculated using the Aperture Photometry Tool software \citep{APT} with uncertainties conservatively estimated at 10 per cent.

\begin{table*}
\caption{VES~735 campaign observations and measured \halph line parameters. See Section~\ref{sec:obs} for full description of columns.}
    \label{tbl:obs}
    \centering
    \begin{tabular}{lcccccccc} 
    \hline
Obs.	& Epoch & Date (UTC) &	MJD	&	EW	&	V/R	&	$\Delta$RV	&	E/C & W	\\
&&&&(\r{A})&& (\kms) &&(\r{A})\\
\hline
\hline
DDO\textsuperscript{1}	& I & 1996-10-15T00:00:00.0 &	50371	&	-1.25	&	1.01	&	467	&	1.20	&  19.8 \\
\hline
DDO\textsuperscript{1}	&I& 1997-01-15T00:00:00.0 &	50463	&	-2.76	&	1.09	&	395	&	1.24    &	20.0 \\
\hline
DAO	&I& 1997-09-20T23:24:43.2 &	50711.9755 &	-2.77	&	0.98	&	370	&	1.19&	19.9\\
\hline
DDO\textsuperscript{1}	&I&	1997-12-15T00:00:00.0 &	50797	&	-1.81	&	1.0	&	401	&	1.41&	    19.0\\
\hline
WIRO	&I&2013-08-29T11:13:30.144&	56533.46771	&	-0.87	&	1.04	&	495	&	1.15&22.0	\\
\hline
WIRO	&I& 	2013-11-30T02:32:19.392 &	56626.10578	&	-2.26	&	1.02	&	423	&	1.21&22.8	\\
\hline
WIRO	&I& 2014-09-01T10:31:04.800 &	56901.43825	&	-0.21	&	0.99	&	529	&	1.08&22.3	\\
\hline
GNIRS\textsuperscript{2}	&II& 2016-10-16T14:26:18.024 & 57677.6015975	& -3.06	& -	& -	& 1.10	& 61.6	\\
\hline
APO	&III& 2019-10-30T05:45:41.184 &	58786.24006	&	-14.14	&	1.07	&	186	&	2.16&32.4	\\
\hline
SOFIA (11.2$\micron$)\textsuperscript{3} & III& 2021-06-30T06:27:36.20 &	59395.26916898	&-	&-	&-	&-	&-	\\
SOFIA (7.7$\micron$)\textsuperscript{3} & III& 2021-07-08T09:30:51.74 &	59403.39643218	&-	&-	&-	&-	&-	\\
\hline
DAO	&III& 2021-11-17T07:11:20.256 &	59535.29954	&	-23.54	&	1.54	&	178	&	4.34	&	19.1	\\
DAO	&III& 2021-11-17T07:41:25.152 &	59535.32043	&	-23.41	&	1.54	&	180	&	4.42	&	18.1	\\
\textbf{DAO}	&\textbf{III}&	\textbf{2021-11-17T07:26:22.704} &	\textbf{59535.309985} &	\textbf{-23.48}	&	\textbf{1.54}	&	\textbf{178}	&	\textbf{4.38} & \textbf{18.6}	\\
\hline
DAO	&III& 2021-11-24T05:53:08.736 &	59542.24524	&	-23.44	&	1.54	&	197	&	4.48	&	20.2	\\
DAO	&III& 2021-11-24T06:23:15.360 &	59542.26615	&	-23.24	&	1.57	&	192	&	4.47	&	19.9	\\
\textbf{DAO}	&\textbf{III}&	\textbf{2021-11-24T06:08:12.048} &	\textbf{59542.255695}	&	\textbf{-23.34}	&	\textbf{1.55}	&	\textbf{192}	&	\textbf{4.50 }& \textbf{20.0}	\\
\hline
DAO	&III&	2021-12-19T05:42:45.792 &	59567.23803	&	-22.74	&	1.44	&	176	&	4.40	&	19.5	\\
DAO	&III& 2021-12-19T06:12:52.416 &	59567.25894	&	-22.20	&	1.47	&	182	&	4.32	&	20.5	\\
\textbf{DAO	}&\textbf{III}& \textbf{2021-12-19T05:57:49.104 }&	\textbf{59567.248485}	&	\textbf{-22.47	}&	\textbf{1.44}	&	\textbf{177	}&	\textbf{4.33} & \textbf{20.0}	\\
\hline
DAO	&III& 2021-12-21T03:45:21.600 &	59569.1565	&	-22.35	&	1.46	&	182	&	4.33	&	19.5	\\
DAO	&III& 2021-12-21T04:15:27.360 &	59569.1774	&	-22.30	&	1.47	&	175	&	4.32	&	18.8	\\
DAO	&III&	2021-12-21T04:45:33.120 &	59569.1983	&	-22.50	&	1.45	&	180	&	4.32	&	18.8	\\
\textbf{DAO}	&\textbf{III}&	\textbf{2021-12-21T04:15:27.360} &	\textbf{59569.1774}	&	\textbf{-22.38}	&\textbf{	1.45}	&	\textbf{178	}&	\textbf{4.35} & \textbf{19.0}\\
\hline
APO	&IV&	2022-05-29T10:42:31.680 &	59728.4462	&	-17.20	&	0.76	&	213	&	4.14&16.2	\\
\hline
DAO	&IV& 2022-10-15T08:32:47.040 &	59867.35610	&	-20.72	&	0.69	&	193	&	4.52	&	18.8	\\
DAO	&IV&	2022-10-15T09:03:31.680 &	59867.37745	&	-20.68	&	0.67	&	188	&	4.47	&	19.1	\\
DAO	&IV&	2022-10-15T09:33:38.304 &	59867.39836	&	-20.68	&	0.69	&	201	&	4.50	&	19.2	\\
DAO	&IV&	2022-10-15T10:03:43.200 &	59867.41925	&	-20.40	&	0.66	&	188	&	4.49	&	19.3	\\
DAO	&IV&	2022-10-15T10:33:48.960 &	59867.44015	&	-20.70	&	0.68	&	189	&	4.31	&	20.3	\\
DAO	&IV&  2022-10-15T11:03:54.720 &	59867.46105	&	-20.44	&	0.68	&	181	&	4.42	&	18.9	\\
\textbf{DAO}	&\textbf{IV}&	\textbf{2022-10-15T09:48:34.272} &	\textbf{59867.40873}	&	\textbf{-20.60}	&	\textbf{0.69}	&	\textbf{196}	&	\textbf{4.44}& \textbf{19.3}	\\
\hline
DAO	&IV&	2023-01-11T03:09:07.776 &	59955.13134	&	-23.28	&	0.80	&	172	&	4.51	&	18.5	\\
DAO	&IV& 	2023-01-11T03:39:12.672 &	59955.15223	&	-22.60	&	0.76	&	171	&	4.47	&	19.0	\\
DAO	&IV&	2023-01-11T04:09:19.296 &	59955.17314	&	-22.04	&	0.75	&	178	&	4.68	&	19.2	\\
\textbf{DAO}	&\textbf{IV}&	\textbf{2023-01-11T03:39:13.536} &	\textbf{59955.15224}	&	\textbf{-22.64}	&	\textbf{0.79}	&	\textbf{173}	&	\textbf{4.58} & \textbf{18.9}	\\
         \hline
         \hline
         \multicolumn{9}{l}{\textsuperscript{1} The precise dates of these observations are not know and have been estimated to the 15th of the month in which we know the }\\
         \multicolumn{9}{l}{observation occurred.}\\
         \multicolumn{9}{l}{\textsuperscript{2} Note that GNIRS data are measured parameters of single-peaked Br$\gamma$ emission rather than \halphend.}\\
         \multicolumn{9}{l}{\textsuperscript{3} SOFIA photometry are found in Table \ref{tbl:phot}.}\\
    \end{tabular}  
\end{table*}

We obtained archival $V$-band photometry for VES~735 from the All-Sky Automated Survey for Supernovae \citep[ASAS-SN;][]{Shappee,Kochanek}, and $g$-band photometry from the Zwicky Transient Facility \citep[ZTF;][]{Masci}. Additional, primarily $V$-band, photometry was obtained by American Association of Variable Star Observers (AAVSO) volunteers participating in an observing campaign supporting our DAO spectroscopic observations in late 2022. The start and end of the $g$-band data from ZTF overlapped in time with the  ASAS-SN and AAVSO coverage respectively. A $-0.82$ magnitude offset to the ZTF photometry was applied to bring it into agreement with both the ASAS-SN and AAVSO photometry.  For simplicity, we refer to this offset $g$-band data as $V$-band for the remainder of the paper.

We retrieved archival \textit{Spitzer}, \textit{WISE} and \textit{NEOWISE} photometry from IRSA. \textit{Spitzer} photometry at 4.5, 5.8, and 8.0 $\mu$m was obtained using the \textit{Spitzer} Enhanced Image Products (SEIP) source list (SSTSL2), and 3.6 $\micron$ photometry was obtained from the GLIMPSE360 catalog. \textit{WISE} surveyed the entire sky at 3.4, 4.6, 12, and 22~$\micron$ to 5$\sigma$ sensitivity of roughly 16.6, 15.6, 11.3, and 8.0 magnitudes respectively \citep{Wright} and was put into hibernation in early 2011. In late 2013, \textit{WISE} was brought out of hibernation and re-purposed as \textit{NEOWISE}, operating at only the 3.4 and 4.6~$\micron$ bands \citep{NeoWise}. Using the same observational cadence as the original mission, \textit{NEOWISE} covers the nearly the entire sky once every six months with a minimum of eight independent exposures (\app 12 exposures on average) with the same level of photometric quality seen during the pre-hibernation mission \citep{NEOWISE14}.

\begin{table}
\caption{Infrared photometry used for modelling VES~735.}
    \label{tbl:phot}
    \centering
    \begin{tabular}{lcccc} 
    \hline
Telescope & Band & Epoch & F($\nu$) & eF($\nu$)  \\
&&Observed&(mJy)&(mJy)\\
\hline
\hline
\textit{2MASS} & J & I & 331.5 & 7.6 \\
\textit{2MASS} & H & I & 361.7 & 11.3 \\
\textit{2MASS} & K\textsubscript{S} & I & 328.6 & 8.2 \\
\textit{Spitzer} & I1 & I & 142.0 & 5.6\\
\textit{Spitzer} & I2 & I & 102.90 & 0.035 \\
\textit{Spitzer} & I3 & I & 87.75 & 0.072 \\
\textit{Spitzer} & I4 & I & 58.99 & 0.040\\
\textit{WISE} & W1 & I & 164.3 & 3.5 \\
\textit{WISE} & W2 & I & 108.4 & 2.1 \\
\textit{WISE} & W3 & I & 43.1 & 1.0 \\
\textit{NEOWISE}\textsuperscript{1} & W1 & III & 830.8 & 54.1 \\
\textit{NEOWISE}\textsuperscript{1} & W2 & III & 456.2 & 16.0 \\
\textit{NEOWISE}\textsuperscript{2} & W1 & III & 528.1 & 41.0 \\
\textit{NEOWISE}\textsuperscript{2} & W2 & III & 380.3 & 7.7 \\
\textit{SOFIA} & 7.7$\micron$ & III & 197 & 20 \\
\textit{SOFIA} & 11.2$\micron$ & III & 134 & 13 \\
         \hline
         \hline
        \multicolumn{5}{l}{\textsuperscript{1}\textit{NEOWISE} values and uncertainties were determined from the}\\
        \multicolumn{5}{l}{average and standard deviation of the observations taken from }\\
        \multicolumn{5}{l}{the MJD 59236 data release.}\\
        \multicolumn{5}{l}{\textsuperscript{2} Same as in \textsuperscript{1} but for the MJD 59444 data release.}\\
    \end{tabular}  
\end{table}

\begin{figure}
    \centering
    \hspace*{-1.cm}
    \includegraphics[scale=0.45]{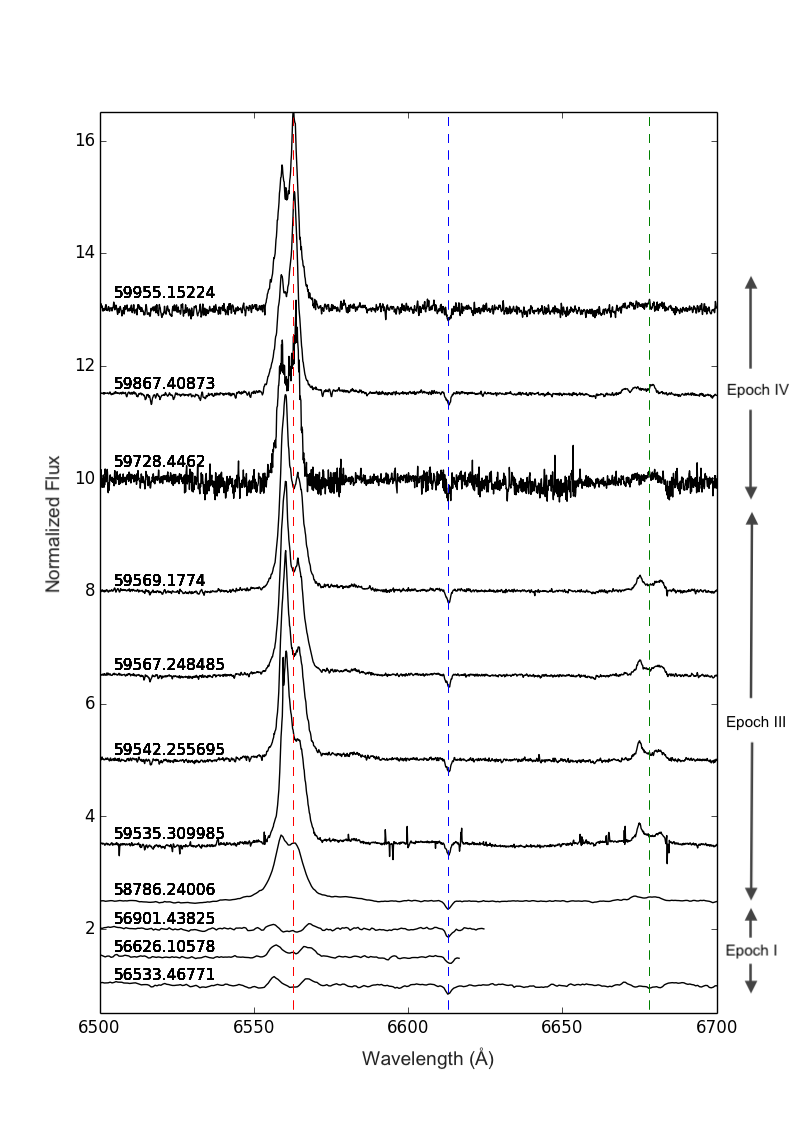}
    \caption{Normalized \halph \& \he spectra of VES~735 from 2013 -- 2023. The line intensities in each spectra have been normalized to the stellar continuum. The red, blue, and green dashed lines are centered on the rest wavelengths of \halph, the $\lambda$6613 \r{A} DIB, and \he, respectively}
    \label{fig:stack}
\end{figure}

\begin{figure*}
    \centering
    \includegraphics[width = \textwidth]{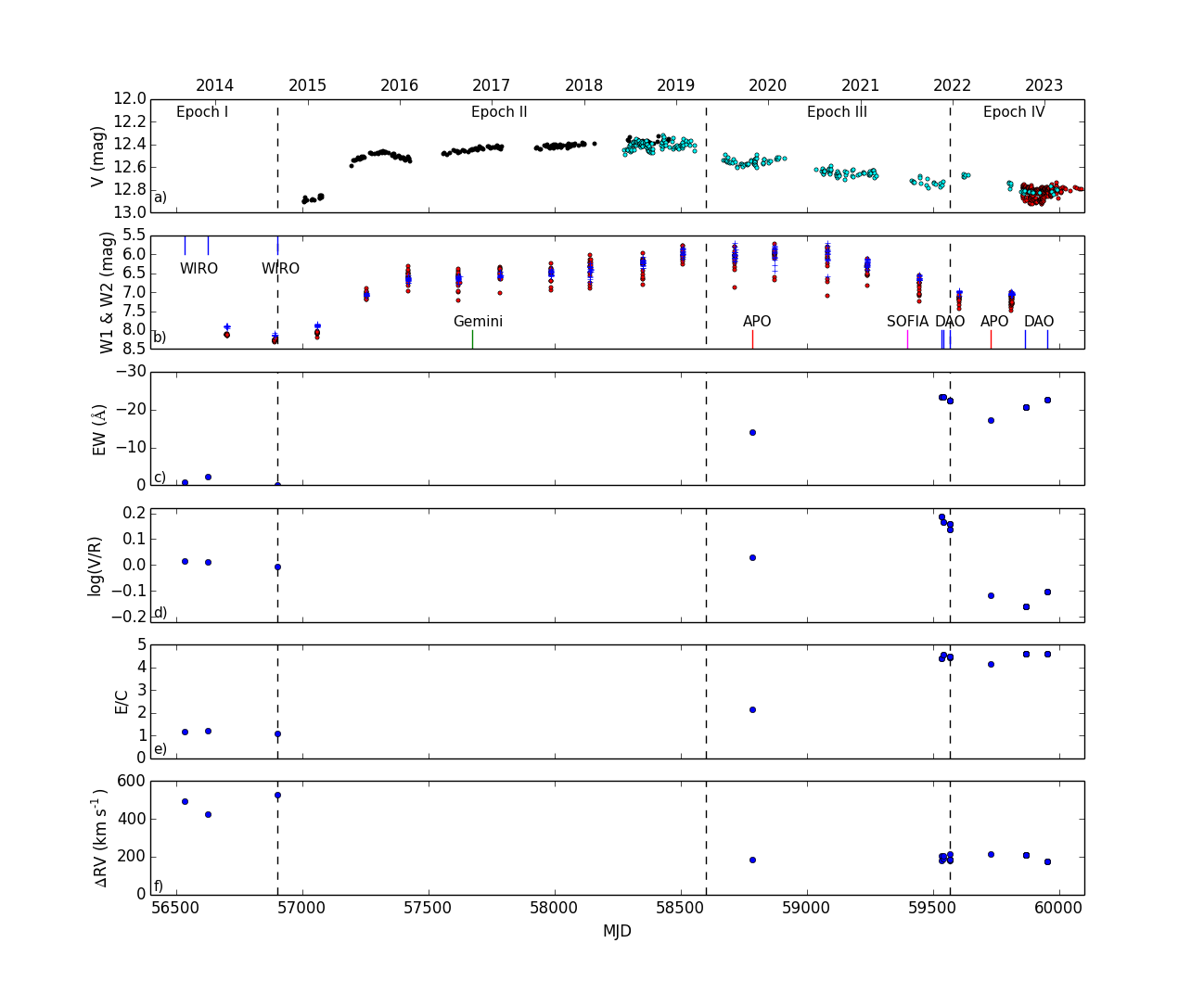}
    \caption{Time variability of VES~735 brightness and parameters of the \halph line. Panel a) displays the $V$-band magnitude (ASAS-SN shown in black, ZTF shown in cyan, AAVSO shown in red), b) the \textit{NEOWISE} infrared magnitudes (W1 shown in blue, W2 shown in orange) c) the Equivalent Width in \r{A}, d) the log(V/R) peak intensity ratio, e) the peak intensity normalized to the continuum, and f) the separation of the double peak in \kms. The dashed lines separate the epochs described throughout Section~\ref{sec:results}. For reference, the date of other notable observations are displayed in panel b)}.
    \label{fig:meas}
\end{figure*}

VES~735 \halph spectra, separated by epoch, are shown in Fig.~\ref{fig:stack}. Photometry and measured \halph line parameters are shown as a function of MJD in Fig.~\ref{fig:meas}.

\section{Spectral Type Justification}
\label{sec:spectype}

Spectral classification of Oe stars is challenging due to the possible infilling of the photospheric lines typically used for classification from circumstellar emission. The main concern is that the \ion{He}{i} line used will appear too weak in comparison to the \ion{He}{ii} line, resulting in too early a spectral classification (e.g., see \citealt{neg2004}). \citet{KBM99} determined the spectral type of VES~735 by constructing an atlas of OB stars in the 4800--5420~\r{A} region and calibrating spectral type with the equivalent width ratio of the \hei and \heii lines, resulting in a spectral class of O8.5$\pm0.5$. The luminosity class (V) was determined from optical ($B$ and $V$) photometry combined with knowledge of the distance to the associated \hii region. They examined the possible effect of \hei line infill by calibrating spectral type versus the equivalent width of the individual \hei and \heii lines and found spectral types of O8.5 and O7.5, respectively, with uncertainties of $\pm1$ spectral type. Based on this consistency, they concluded that in-filling of the \hei line did not play a significant factor in the determination of the spectral type.

\cite{KBM99} also used 1420~MHz radio continuum observations of the surrounding \ion{H}{ii} region KR~140 to determine the spectral type of VES~735 by comparing the derived Lyman continuum luminosity ($Q$), in photons s$^{-1}$, with that predicted from various stellar atmospheric models. 
They found $\log(fQ) = 48.09\pm0.11$, using a distance of $2.3\pm0.3$~kpc based on the potential association with the nearby IC~1805 star cluster, and estimated a covering factor of $f = 0.4 - 0.5$, consistent with the observed blister morphology of the \ion{H}{ii} region. We rescaled their analysis using the new \textit{Gaia} \citep{Gaia} distance estimate (EDR3) to VES~735 of \app 1.45 kpc and found $\log(Q)$ lies between 47.99 and 48.09 for the same range of covering factor values. This Lyman continuum luminosity range still corresponds to late-type O stars for most atmospheric models (\citet{Panagia73}: O9.5~V, \citet{Thompson84}: O9~V, \citet{Martins05}: O8.5~V -- O9~V, \citet{Crowther05}: O9~V -- O9.5~V). 

\begin{figure} 
    \centering
    \includegraphics[scale=0.4]{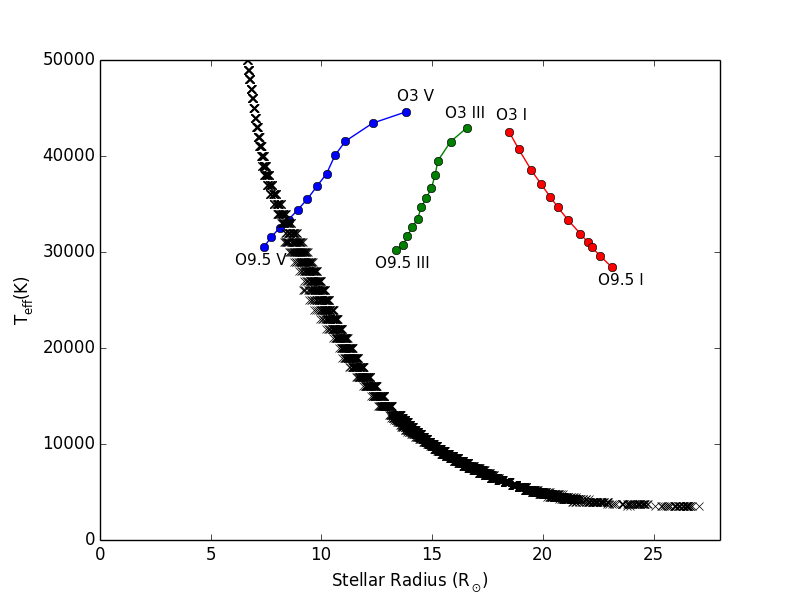}
    \caption{Spectral type determination of VES~735 using SED fitting of stellar atmosphere models from \citet{kurucz93}. The best-fit SED models for VES~735 are shown as black crosses. The stellar calibrations from \citet{Martins05} are shown as blue, green, and red points for spectral types O3, O4, and O5 -- O9.5 (at intervals of 0.5 in spectral type) for luminosity class V, III, and I respectively.}
    \label{fig:ostar}
\end{figure}

As an independent check on the spectral type, we used a classification method adapted from \citet{Watson08}. $J$, $H$, and $K_s$-band photometry from the Two-Micron All Sky Survey \citep[2MASS; ][]{skr06} and 3.6--8.0 $\mu$m photometry from the \emph{Spitzer} Infrared Array Camera \citep[IRAC; ][]{faz2004} were used to construct a spectral energy distribution (SED) for VES~735. The SED was then analysed by the \citet{rob07} SED fitting code using the suite of stellar atmospheres models from \citet{kurucz93}. The SED models determined to be the best fit to VES~735 were output and scaled to its distance. The model  $T$\textsubscript{eff} and stellar radius were then be compared with parameters expected for different stellar calibrations. Fig.~\ref{fig:ostar} shows the locus of the best-fit models (black crosses) intersects with the locus of O-star $T$\textsubscript{eff}-R relations of \citet{Martins05} (blue, green, and red points for luminosity classes V, III, and I, respectively) at \app O8.5~V further supporting the original spectral classification. We adopt O8.5~Vpe as the spectral type for VES~735.

\section{Results}
\label{sec:results}

\subsection{Epoch I: Long-Lived, Double-Peaked \halph}
\label{sec:EpochI}

VES~735 was catalogued in the emission-line survey by \citet{VAT}; however, the double-peaked nature of the \halph line was not known until it was first observed at DDO in 1996. \citet{KBM99} observed VES~735 three further times in 1997 to monitor the longevity and variability of the emission (see their figure 8). Relatively little variation in the line profile was observed, and they determined a typical equivalent width of $-2.5$~\r{A}  and peak separation of \app 400 \kms. Photometry obtained at DAO at this time showed a $V$-band magnitude of $12.88\pm0.01$.

\citet{KBM99} extended the spectral coverage of VES~735 further into the red to include the \ion{He}{i} $\lambda$6678 line to see if the absorption feature could be used to further classify the star \citep{RedClass}. However, each observation throughout Epoch I that covered the \ion{He}{i} $\lambda$6678 line showed the absorption feature to be filled in, and in some case slightly above the continuum level (coverage of this line was not included on MJD 56626 and 56901). 

Observations acquired from WIRO showed that the \halph line was still in emission, double-peaked, and had a peak separation ranging between 400 and 550 \kms (see bottom of Fig.~\ref{fig:stack}). The emission was somewhat variable in strength displaying equivalent widths between $-0.2$ and $-2.3$~\r{A}. No additional $V$-band measurements were made during Epoch I. 

In addition to displaying modest, long-lived \halph emission, VES~735 had a slight near- and mid-infrared excess during Epoch I. Following \citet{lada87}, the SED between 2MASS $K_s$ and the IRAC 8.0~$\micron$ band has $\alpha=-2.32$. However, VES~735 is heavily reddened, with A\textsubscript{V} = 5.5 \citep{KBM99}, so we also compute the spectral index from dereddened photometry using extinction law from \citet{indeb}, finding a value of $\alpha=-2.57$. 

To estimate the expected value of $\alpha$ for a massive star with no circumstellar material, we used model photometry from the \textit{Spitzer} Stellar Performance Estimation Tool (STAR-PET \footnote{\url{https://irsa.ipac.caltech.edu/data/SPITZER/docs/dataanalysistools/tools/pet/starpet}}). STAR-PET uses a user-input K-band magnitude to estimate the IRAC flux densities for stellar point sources of different spectral types based on Kurucz-Lejeune atmospheric models \citep{kurucz93, lejeune97}. For an O8~V star (the nearest available to an O8.5~V spectral type) the expected value of the SED slope is $\alpha=-2.98$.  We also computed $\alpha$ for five other O-type main-sequence stars with similar spectral types and having coverage from 2MASS K\textsubscript{S} through IRAC 8.0$\micron$ (see Table~\ref{tab:spectralIndex}). We see that, while the observed, dereddened, $\alpha$ value of VES~735 differs from the STAR-PET value by 0.41, all of the other stars differ from the model values by only $\sim 0.10$. 

\begin{table}
\caption{Infrared spectral index comparison of main sequence O-type stars. Spectral indexes were calculated using 2MASS $K_s$ and IRAC 8.0~$\micron$ photometry. Dereddening was done using the \citet{indeb} extinction law and intrinsic stellar colours from \citet{mem23}. Note that the STAR-PET tool has a limited number of spectral types to choose from, so each star listed was modeled as either an O8~V or O6~V, whichever was nearest to its listed spectral type.}
    \label{tab:spectralIndex}
    \centering
    \begin{tabular}{lcccc}
    \hline
        Star    & Spectral      & \multicolumn{3}{c}{Infrared Spectral Index ($\alpha$)}    \\
                & Type          & Observed & Dereddened  &  STAR-PET   \\
        \hline
        \hline
        VES 735   & O8.5Vpe                       & $-2.32$ & $-2.57$  & $-2.98$ \\
        ALS 15111 & O8V C\textsuperscript{a}      & $-2.67$ & $-2.90$  & $-2.98$ \\
        HD 46056  & O8Vn C\textsuperscript{b}     & $-3.01$ & $-3.08$  & $-2.98$ \\
        HD 167633 & O6.5V((f))\textsuperscript{b} & $-2.84$ & $-2.90$  & $-2.98$ \\
        HD 305532 & O6.5V((f))\textsuperscript{c} & $-2.79$ & $-2.88$  & $-2.99$ \\
        HD 344784 & O6.5V((f))\textsuperscript{b} & $-2.73$ & $-2.85$  & $-2.99$ \\
        \hline
        \multicolumn{5}{l}{\textsuperscript{a} \citet{als}, \textsuperscript{b} \citet{Sota11}, \textsuperscript{c} \citet{Sota14}.}
    \end{tabular}
    
\end{table}

\textit{NEOWISE} W1 (3.4~$\mu$m) and W2 (4.6~$\mu$m) photometry is available from two data releases near the end of Epoch I, first on MJD 56702 and again on 56892. The first data release, consisting of 13 individual observations, shows a tight grouping with average ($\pm$ standard deviation) W1 and W2 magnitudes of $8.126\pm0.015$ and $7.906\pm0.014$ respectively (see Fig. \ref{fig:meas}). This matches very well with the \textit{WISE} photometry, which would have been obtained approximately four years earlier: $\mathrm{W1}=8.16\pm0.02$ and $\mathrm{W2}=7.97\pm0.02$. The second data release, consisting of 16 observations, also shows a tight grouping, but both W1 and W2 have each dimmed by \app 0.2 mag to $8.276\pm0.025$ and $8.126\pm0.014$ respectively.

The double-peaked emission and the slight infrared excess are both well-modelled as emission associated with an optically thin disc of circumstellar gas (see Section~\ref{sec:basic} and Section~\ref{sec:irexcess} respectively). The longevity and consistency of the double-peaked emission, along with the association of VES~735 with a young \ion{H}{II} region, led \citet{MK18} to the assumption that VES~735 may be harboring a remnant accretion disk from its formation and could potentially be classified as a `Herbig' Oe star analogous to lower mass young stellar objects. However, due to the recent behavior, which will be discussed in the sections below, we instead find it more likely that this is a remnant \textit{decretion} disk. In this case the material seen in Epoch I is a part of a previously unknown cyclic behavior similar to what has been observed in Epochs II, III, and IV, and the enhanced emission seen in \textit{WISE} and \textit{Spitzer} photometry is due to thermal bremsstrahlung emission from heated gas in the circumstellar environment rather than excess infrared radiation characteristic of a dusty protoplanetary disk.

 \subsection{Epoch II: Photometric Brightening}
 \label{sec:EpochII}

Epoch II is defined as starting after our final WIRO spectra and extending through the ASAS-SN photometric coverage into the first ZTF data release having VES~735 coverage. $V$-band photometry is available from the ASAS-SN survey that extends from MJD 57007 to 58451. Fig.~\ref{fig:meas} shows a sharp rise in the $V$-band from 12.85 to 12.59 occurs between MJD 57078 and 57197. During this same span of time, \textit{NEOWISE} photometry also shows an increase of \app 1 mag in both the W1 and W2 bands. Over the course of the next \app 3.5 years, there is a much more gradual increase to a maximum brightness of $V = 12.35$, when the ASAS-SN coverage ends. The \textit{NEOWISE} photometry mimic this more gradual brightening, with the W1 and W2 bands each showing an additional increase of \app 0.5 mag through the remainder of Epoch II. In contrast to the infrared data in Epoch I, there is also a very noticeable spread in magnitudes during this period of gradual brightening, with variability as high as a full magnitude over the course of hours and days during each \textit{NEOWISE} observation period for both W1 and W2.

The ZTF observations overlap with the last set of ASAS-SN photometry in late 2018, which allows us to offset the ZTF data such that the average ZTF and ASAS-SN magnitudes in this time period match. The ZTF observations begin to show a slow decline in brightness at the end of Epoch II, which continues through to Epoch IV.

We find the \textit{NEOWISE} and $V$-band photometry described above to show a decretion event commonly described by the mass reservoir effect for viscous decretion discs (VDDs) around Be stars \citep{ghoreyshi2017,rimulo2018}. Through some mechanism, material is being lifted into a circumstellar disc. \citet{Riv13} explains that this disc can be thought of as a \textit{pseudo-photosphere}, which often causes a net brightening depending on the size, density, and inclination angle of the disc. For edge-on, low-density systems, the flow of material will cause a net dimming effect, which allows us to rule out these parameters for VES~735 (see Section \ref{sec:basic}). Their models go on to show that for a disc at an inclination angle $i = 30^{\circ}$, 80 per cent of the V-band flux originate within just a couple stellar radii whereas the same fraction of near- to mid- infrared flux comes from from a larger volume of the disc (see their section 2.1). Thus, the initial flow of material will cause a sharp rise at visible and infrared magnitudes due to the increase in density of the inner portion of the disc. Assuming that the disc is being fed at a constant rate for a long period of time, the inner disc can reach a relatively stable configuration where it is still being fed without much additional contribution to the $V$-band emission. This continual feeding will begin to supply the outer portions of the disc with material, causing infrared and radio emission to continue to grow even after the $V$-band has plateaued. Fig.~\ref{fig:meas} shows precisely this effect during Epoch II where there is a sharp rise in the $V$-band magnitude in just a few months, followed by a few years of relative stability while the infrared emission continues to rise. 

\begin{figure}
    \label{fig:kbandspec}
    \centering
    \includegraphics[scale=0.4]{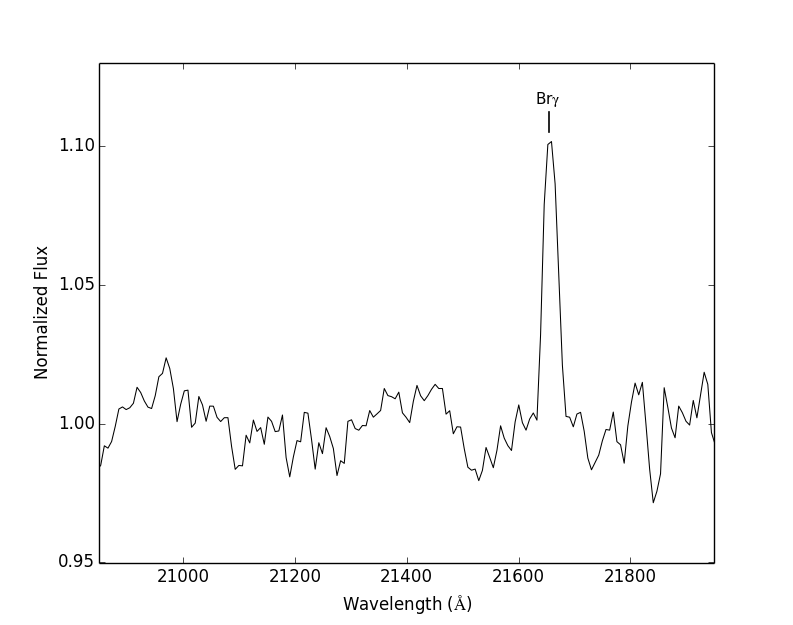}
    \caption{Normalized $K$-band spectrum (\app 12~\r{A} resolution) of VES~735. The spectrum was obtained on MJD 57677 using the Gemini Near-Infrared Spectrometer (GNIRS) on the 8-m Gemini North telescope.}
    \label{fig:gemini}
\end{figure}

The only spectra obtained during Epoch II was taken in the $K$-band by GNIRS. The aim was to search for indicators of a circumstellar disc such as CO bandheads and Br$\gamma$ emission. The former is expected to be visible if there is a dense disc, and the latter can arise from a disc-wind formed from heating the surface of a dense disc by intense ultraviolet radiation from the central star \citep{bik}. VES~735 did show Br$\gamma$ in emission during this observation with a FWHM = 405 $\pm$ 15 \kms (see Fig.~\ref{fig:gemini}). Such a large width suggests emission from a circumstellar rotating disc rather than the surrounding nebula. Br$\gamma$ emission is also thought to be a tracer for magnetospheric accretion, however the correlations seen from spectro-interferometry studies of Herbig AeBe stars by \citet{kraus2008} also favour a stellar- or disc-wind interpretation for their sample of stars, particularly those which also exhibit double-peaked \halph emission. Because this is the only $K$-band spectrum available, it is unclear whether this line was in emission prior to Epoch II, or whether this occurred in concert with the increase of the $V$-band brightness.

\subsection{Epoch III: Spectroscopic Infall Signatures} 
\label{sec:EpochIII}

\begin{figure}
    \centering
    \includegraphics[scale=0.4]{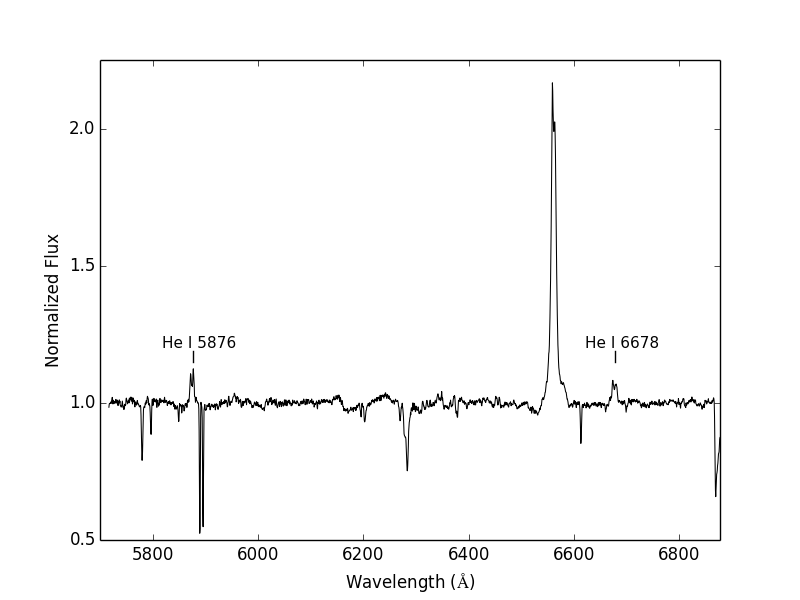}
    \caption{Continuum normalised spectra of VES~735 from the Apache Point Observatory on MJD 58786 from 5700 -- 6877\r{A}.}
   \label{fig:apo}
\end{figure}

Epoch III is defined by the steady decline in the ZTF $V$-band brightness of VES~735, starting from the peak in Epoch II and gradually decreasing by about 0.3 mag over the next two years (see Fig.~\ref{fig:meas}). The \textit{NEOWISE} W1 and W2 band photometry both plateau at \app 6.0 mag in the middle of Epoch III, before starting to decrease. Similar to Epoch II, the W1 and W2 magnitudes show large variability ($>1.0$ mag in some cases) over the course of hours in each \textit{NEOWISE }data release.

Photometric observations from the \textit{SOFIA} telescope were taken at 7.7, 11.2, 25.3, and 34.8 $\micron$; however, VES~735 was detected only at 7.7 and 11.2 $\micron$. Interestingly, when comparing the \textit{SOFIA} photometry at these wavelengths with archival IRAC 8 $\micron$ and \textit{WISE} 12 $\micron$ data, respectively, the \textit{SOFIA} flux densities are greater than the archival data by a factor of three in both cases (see Table \ref{tbl:phot}). This is consistent with the increase in brightness seen at shorter infrared wavelengths in the \textit{NEOWISE} data.

The photometry throughout Epoch III again show a consequence of the mass reservoir effect for VDDs. \citet{rimulo2018} describe how that after mass-injection has ceased, the inner disc will remain dense when re-accretion occurs and the cessation of the mass-injection will not be reflected in the $V$-band light curves for a period of time. Following this interpretation, we would assume the mass reservoir for VES~735 was fed for \app 3 years as seen by the plateau in the $V$-band followed by disc dissipation/re-accretion beginning at some point between 2018 and 2019. The longer that mass-injection takes place, the longer the disc will take to dissipate when re-accretion begins. As the reservoir is now feeding back to the inner disc, we see a slow $V$-band decline from 2019 to 2023. As discussed by \citet{carciofi2012}, the re-accretion of material is caused by turbulent viscosity in the inner disc transferring angular momentum to the outer portions of the disc. While the inner disc material travels back inward due the loss of angular momentum, material can still be fed to the outer disc. This effect is shown by the \textit{NEOWISE} light curves as they continue to grow well into Epoch III after mass-injection from the star to the inner disc has ceased. 

\begin{figure}
    \centering
    \hspace*{-0.6cm}
    \includegraphics[scale = 0.45]{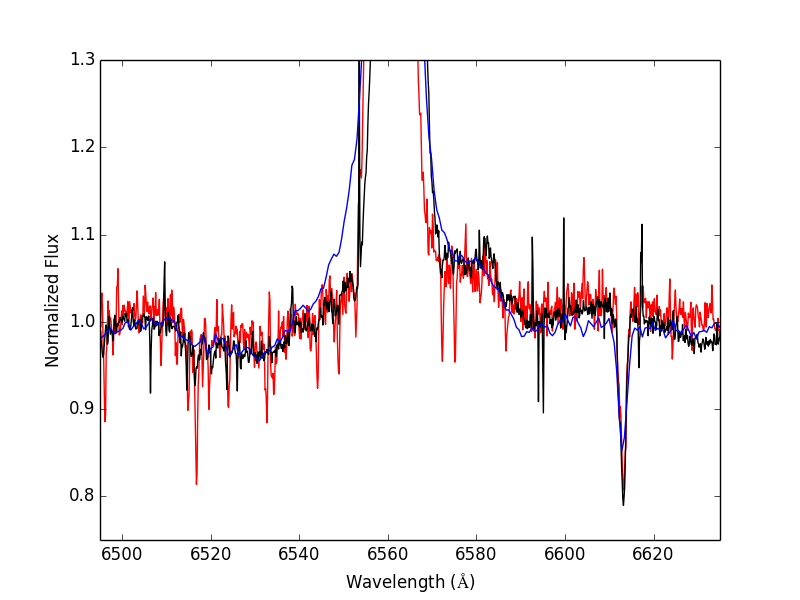}
    \caption{Continuum normalised spectra of VES~735 from 6495 -- 6635\r{A} to highlight the P~Cygni profile and red emission shelf. The blue line is the from MJD 58786.24006 from the APO, red shows the average combined spectra at MJD 59867.40873 from DAO (see Table~\ref{tbl:obs}), and black shows the average combined spectra between MJD 59535.29954 and 59569.1983 from the DAO.}
    \label{fig:wind}
\end{figure}

In contrast with the decreasing $V$-band photometry, the \halph line has grown to show an equivalent width of $EW = -14.14$~\r{A} and a full-width at the continuum level of $W=32.4$~\r{A} as seen by the APO spectrum in Fig. \ref{fig:apo}; both are a dramatic increase compared to Epoch I values (see the lower part of Fig.~\ref{fig:stack}). In addition, the \halph line shows a P~Cygni profile extending blueward to \app 6510 \r{A} as well as a red emission ‘shelf’ that is visible to nearly 6590 \r{A} (see Fig. \ref{fig:wind}). Both of these features are associated with a more spherically symmetric outflow, and are more commonly associated with supergiant stars. The P~Cygni profile is famously associated with absorption from an expanding wind \citep{Beals1953}, while the red shelf emission is likely due to electron scattering in an expanding wind \citep{auer1972}. \citet{cl74} show an example of this type of structure in the \halph profile of the O8 If star 9 Sge (see their figure 10). In VES~735, the P~Cygni profile has a depth of 0.965, on the continuum-normalised spectrum, and the red shelf has a height of 1.07. At this point in time, the V/R peak ratio is just slightly above unity, and the peak separation has been reduced to \app 200~\kms. This APO spectrum includes the \ion{He}{i} $\lambda$5876, \halphend, and the \ion{He}{i} $\lambda$6678 lines. Each of these lines is double-peaked, and the centroid of each line is slightly blue-shifted. Further observations only cover the \halph and \ion{He}{i} $\lambda$6678 lines.

The remainder of the spectral observations in Epoch III were collected two years later at DAO. As illustrated in Fig.~\ref{fig:stack}, the \halph profile has continued to increase with the peak intensity more than doubling what was seen at APO two years prior. The EW has increased to \app $-23$ \r{A}, the V/R sits well above unity ranging between 1.48 and 1.54, and the peak separation remains fairly constant around 200 \kms. The rest wavelength for the \halph line is now sitting in the centre of the trough of the double-peak. The P~Cygni profile is a bit more difficult to discern in the DAO spectra since the blue cutoff of the spectra is at the extent of the absorption feature; however, the depth of the feature, $0.98\pm0.01$, appears to be the same as in the APO spectrum, and the red emission shelf has the same height of $1.08\pm0.02$.

These \halph line variations are also consistent with the mass reservoir effect and VDD modeling of Be stars such as $\omega$ CMa described by \citet{ghoreyshi2021}. The first spectroscopic observations from APO show a large increase of \halph emission (see Fig.~\ref{fig:apo} and  Fig.~\ref{fig:meas}) compared to the intensity seen in decades prior, and occurs when the $V$-band brightness has already been declining. At this point the disc has been fed for years and re-accretion back onto the star may have already begun; however, the \halph line approximately doubles in EW and E/C when observed by the DAO 2 years later at the end of the phase. 

The explanation for this is likely a combination of two factors. First, as described previously, the outer disc is still able to expand even after re-accretion has started. This is clearly seen as the infrared brightness is continuing to rise into Epoch III. As \halph traces a large volume of the disc, the EW should continue to rise until re-accretion/dissipation of the outer disc begins, and the infrared brightness begins to drop. Second, since the increase in brightness is not limited to the $V$-band but extends all the way out to the infrared, this shows that the overall continuum emission of VES~735 has risen (see \textit{NEOWISE} and \textit{SOFIA }photometry in Table~\ref{tbl:phot}), which results in both the EW and E/C decreasing. We see the largest EW and E/C at the end of Epoch III as the visible and infrared magnitudes are approaching their pre-outburst brightness, lowering the overall continuum level. \citep{Clark2001, ghoreyshi2021}

In addition to the overall increase in strength of the \halph lines observed at DAO at the end of Epoch III, these profiles also display an overall blueward shift and a strong asymmetry with V/R ratios of \app 1.5, very reminiscent to what is seen in the Oe star AzV~493 \citep{oey2023}. Such \halph profiles are interpreted as resulting from infalling material \citep{hartmann2016}. This is again in agreement with what is seen in the light curves where the infrared light curves have been decreasing for at least one year, suggesting that the outer portions of the disc have stopped being fed and material is now moving back towards the star. Consistent with this picture, the DAO observations show a decrease in the EW of the \halph profile of about 1\r{A} from MJD 59353 to 59569, and this trend continues to the first observation from in Epoch IV on MJD 59728, where the EW has dropped by another 5\r{A}. 

\begin{figure}
    \centering
    \hspace*{-.5cm}
    \includegraphics[scale =0.44]{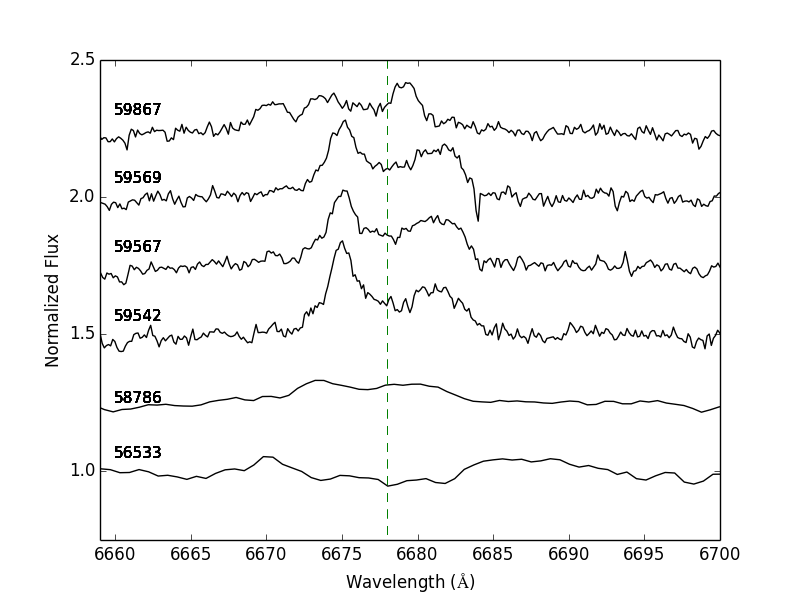}
    \caption{Normalized VES~735 spectra of selective observations between 2013 and 2022. The line intensities in each spectra have been normalized to the stellar continuum and the green dashed line is centered on the rest wavelength of \he.}
    \label{fig:helium}
\end{figure}

\begin{table}
\caption{\he observations of VES~735.}
    \label{tbl:He-obs}
    \centering
    \begin{tabular}{lcccccc} 
    \hline
Obs.	&	MJD	&	EW	&	V/R	&	$\Delta$RV	&	E/C & W	\\
&&(\r{A})&& (\kms) &&(\r{A})\\
\hline
\hline
APO	&	58786	&	-0.88	&	1.01	&	264	&	1.08	&  19.0 \\
DAO	&	59542	&	-1.79	&	1.14	&	275	&	1.33    &	13.7 \\
DAO	&	59567	&	-1.79	&	1.08	&	282	&	1.28&	12.9\\
DAO	&	59569	&	-1.80	&	1.08	&	281	&	1.28&	    13.7\\
DAO	&	59867	&	-1.39	&	--	&	--	&	1.21& 18.0	\\
\hline
\hline
    \end{tabular}  
   
\end{table}

The \he mimics the \halph profile, but on a smaller scale (see Fig. \ref{fig:stack}). In Epoch I the helium line was simply filled in, but at the start of Epoch III, it clearly shows double-peaked emission, and the EW increases until the beginning of Epoch IV. Comparing the APO and DAO observations, respectively (and sequentially), the trough of the double peak moves from a slight blueshift to the rest wavelength, the peak separation remains fairly constant at \app 300 \kms, and the V/R ratio goes from very near unity to as high as 1.14. We believe that these profile changes can be explained via the same mechanisms that cause the \halph variations. The \halph lines form in a large portion of the disc, but the kinematic properties of the helium lines (peak separation and line width) suggest that these form closer to the stellar surface \citep{Clark2001, Riv13}. 

\subsection{Epoch IV: Outflow Signatures?}
\label{sec:EpochIV}

Photometric data at from ZTF shows that the $V$-band magnitude is \app 12.67 at the start of Epoch IV, slightly higher than at the end of Epoch III. The ZTF observations decrease in brightness and overlap with the extensive coverage from AAVSO contributors at the end of Epoch IV. By the end of our observations the $V$-band magnitude has plateaued around 12.8 mag, almost back to its Epoch I brightness (see Fig.~\ref{fig:meas}).

\textit{NEOWISE} photometry shows that the steady decrease in near-infrared brightness seen in Epoch III has levelled out about one magnitude brighter than the Epoch I brightness; however, data are not available after MJD 59813. The infrared variability is also reduced compared to its peak in Epoch III, with the largest $\Delta$mag between consecutive observations for either band being \app 0.5 mag. This is still much larger than what was observed in Epoch I and the very start of Epoch II.

Epoch IV is characterised by a change in the \halph V/R ratio to values much lower than unity for the first time. A spectrum taken \app 160 days after the end of Epoch III has V/R = 0.75, with the red peak of the \halph profile now located at the rest wavelength. DAO spectra show that the V/R ratio remains essentially unchanged throughout Epoch IV. The S/N ratio of the MJD 59569 and 59955 spectra are too low to detect the P~Cygni absorption and red shelf emission features seen in Epoch III; however, these features are seen in the MJD 59867 spectra. Fig. \ref{fig:wind} shows the depth of the absorption is similar to Epoch III ($0.97\pm0.01$) while the emission shelf has decreased slightly in amplitude ($1.06\pm0.01$). The \halph EW has decreased to $-17.2$ \r{A} at the start of Epoch IV, but increases back to $-22.6$ \r{A} in our final observations. The peak separation still remains constant, around 200\kms, throughout Epoch IV.

These profile characteristics are more consistent with outflow events \citep{hartmann2016}, and bear a striking resemblance to profiles generated for an equatorial disc model with high wind velocities by \citet{marl97} to model the emission from the Be star $\psi$ Persei. This explanation would be consistent with the increasing \halph EW from $-17.2$ to $-22.64$ \r{A} and the sharp rise in the $V$-band seen at the beginning of Epoch IV -- the re-accretion may have been interrupted by a new mass-injection event. If new material has been introduced to the disc, this may in turn explain the plateau of the infrared light curve at the end of the \textit{NEOWISE} coverage.

The \he line remains in emission in Epoch IV and continues to mirror the overall changes seen at \halph -- the V/R ratio flip, the infall/outflow profile signatures, and the reduction in EW seen between Epochs III and IV. Additionally, the MJD 59867 observation not only shows that the entire line has shifted blueward, but the profile now displays a third peak and the full width of the line has increased by nearly 5\r{A} (see Fig.~\ref{fig:helium} and Table~\ref{tbl:He-obs}). Again, this corresponds in time to the slight enhancement in the $V$-band brightness and the infrared brightness starting to plateau, both possible indications that another mass-injection event has occurred. 

We caution that the appearance of this third peak is near the noise level of the observation, however if real, this type of V/R variability and appearance of an additional emission peak are sometimes seen in Be stars \citep{Clark2001,Stefl07,Paul17} and are often attributed to the precession of a one-armed spiral density wave within the Be disc. \citet{1991PASJ...43...75O} suggest that the Balmer profile variations will lag behind those lines emitted at the inner portion of the disc. Observations of X Persei from \citet{Clark2001}, although a known binary system, show three and four peaks of the \he line developing over time, with no additional peaks appearing at \halph (see their figures 2 -- 5). If this is the explanation for the variation seen in the \he profile, it may not be surprising that the \halph emission remains double-peaked at the time of the observation. Additionally, for non-binary systems, expected periodic V/R variability range from years to a decade \citep{Okazaki97}. Unfortunately, our spectroscopic coverage of VES~735 is far too sparse to comment on any periodic V/R behavior due to precession within the disc. 

\section{Discussion}
\label{sec:discuss}

\subsection{Binarity}
The \halph emission seen in \citet{KBM99} and \citet{MK18} have profiles that are fairy symmetric about the \halph rest line and show no short-period variations (see Figures \ref{fig:stack} \& \ref{fig:meas}). The first APO spectra on MJD 58786 show a noticeable blueshift at \halphend, which raises the question whether we are seeing the effect of a long-period binary companion. As it happens, the \halph profile remains blueshifted in every subsequent spectrum, though the MJD 58786 shift appears to be the largest of all our observations.

\begin{figure}
    \centering
    \hspace*{-0.6cm}
    \includegraphics[scale = 0.45]{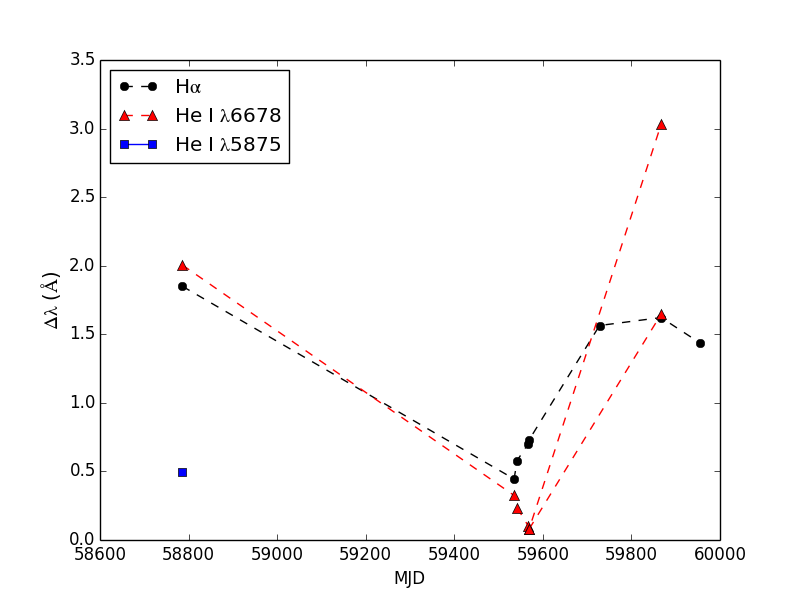}
    \hspace*{-0.6cm}
    \includegraphics[scale = 0.45]{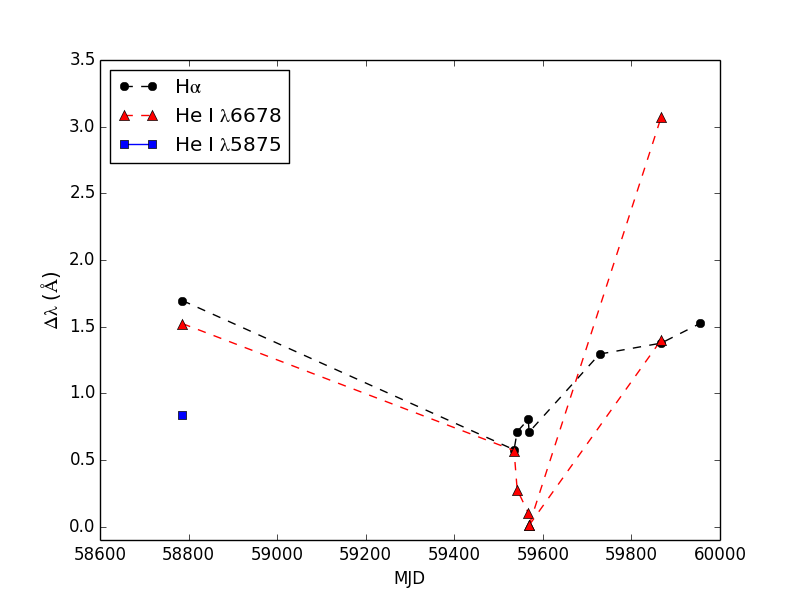}
    \caption{Top Panel: Intensity weighted centroid shifts of \halph and helium lines of VES~735. $\Delta$$\lambda$ represents the shift of the line from their respective rest wavelengths. Positive values of $\Delta$$\lambda$ correspond to an overall blueshift of the line profile. Note that \hel is only covered in the spectrum on MJD 58786 and \he has two measurements for one observation on MJD 59867. The red triangle at the top was calculated including all three observable peaks, the red triangle below only includes the two peaks nearest to the rest wavelength. Measurements were not made for the \he lines for the observations on MJD 59728 and 59955 due to low S/N. Bottom Panel: Same as the top panel except the centroids were determined by Gaussian fitting the outer portions of the emission lines.}
    \label{fig:shift}
\end{figure}

To quantify this, we calculated an intensity weighted central wavelength for the \halphend, \heend, and \hel (only covered once on MJD 58786) emission lines. The MJD 58786 helium centroids were determined by starting at 2 per cent above the continuum emission. Future observations at DAO showed higher intensity and were measured starting at 5 per cent above the continuum, and all \halph spectra were measured starting at 20 per cent above the continuum as to avoid contamination from the red emission shelf. Fig.~\ref{fig:shift} shows that while each line in the MJD 58786 spectrum is blueshifted, they are shifted by different amounts. If these shifts were due to a companion, we would expect similar blueshifts for the \halph and \he lines within the same observation, and we would expect the two lines to show the same variations in blueshift with time, which they do not. We repeated the analysis using Gaussian fits to the outer portions of the \halphend, \heend, and $\lambda$5876 emission lines. The resulting blueshift values and patterns are very similar to our line centroid based analysis. 
A noticeable contrast that we see between the two methods is that the difference in blueshift of the two \ion{He}{i} lines in the MJD 58786 spectrum is smaller when performing the the Gaussian fitting. The large difference seen in the top panel of Fig.~\ref{fig:shift} is due to the \hel line having V/R slightly less than unity (0.98) while the \he line has V/R slightly greater than unity (1.01), which skews its intensity weighted centroid blueward. The position determined by the Gaussian fit method is not influenced by the structure of the center of the lines. While we still see a difference in blueshift of approximately 0.7\r{A} between the two helium lines, this is comparable to the difference in blueshift between the \he and \halph lines seen on MJD 59567 and 56569. These differences are reasonable as different lines likely trace different volumes of the disc and also may experience lags in profile variations as mentioned in Section~\ref{sec:EpochIV}.

While the possibility of a long-period binary companion cannot be completely discounted, we do not prefer this as an explanation for the observed shift in the spectral lines. Rather it is possible that the first \halph profile observed in Epoch III is displaying signatures from both inflow from the inner disk and outflow from the outer disk as described in Section \ref{sec:EpochIII}. The remainder of the profiles from Epoch III again show strong resemblance to AzV 493 with blueshifted emission and redshifted absorption, often characterized as a signature of infalling material. Infalling material would also be expected to produce an inverse P Cygni profile, which VES~735 lacks at \halphend. We think it is possible that this signature is being obscured by the red emission shelf (see Fig. \ref{fig:wind}). Also, the inverse P Cygni profile is not prevalent in \halph profile of AzV 493, though it can be seen prominently in other Balmer lines (see figure 8 in \citet{oey2023}). Unfortunately, we do not have spectral coverage other Balmer lines from this Epoch for VES~735.

\subsection{Basic Disc Parameters}
\label{sec:basic}

The Epoch I \halph spectra, shown at the bottom of Fig.~\ref{fig:stack}, or in more detail in figure 1 of \citet{MK18},  can provide some basic information about the orientation and size of the circumstellar \halph emitting region. Assuming the emission is occurring in an optically thin disc of gas in Keplerian motion around the star \citet{huang1972} showed that the resulting double-peaked profile provides information about the inner and outer radius of the disc (\citet{ebb81} also provides a good, more observationally focused, discussion of the model). In the model, the outer-disc radius ($R_\mathrm{out}$) is constrained by one-half the peak separation velocity ($v_1$) and the inner-disc radius ($R_\mathrm{in}$) is constrained by one-half the full-width velocity at the continuum level ($v_2$). For circular orbits, the disc radii and the velocities are related by
\begin{equation}
\label{eqn:disc}
R_\mathrm{out, in} = \frac{G M_*}{\left(v_{1,2}\right)^2}\sin(i)^2,
\end{equation}
where $i$ is the inclination angle of the disc relative to the plane of the sky.

For Epoch I spectra we find $v_1 \sim 200$~\kms and $v_2 \sim 450$~\kms. Assuming a stellar mass of $M_*= 19.82$~M$_\odot$ and a stellar radius of $R_* = 8.11$~R$_\odot$ \citep{Martins05}, for an inclination angle, $i=90$\degr, these velocities correspond to inner and outer radii of $R_\mathrm{in} = 2.3 R_*$ and $R_\mathrm{out} = 12 R_*$. Since VES~735 brightened as material was injected into the circumstellar disc during Epoch II (see Section \ref{sec:EpochII}) it is likely that the disc is not being seen edge-on. A minimum inclination angle $i_\mathrm{min} = 41.2$\degr~can be determined by setting $R_\mathrm{in} = R_*$, in which case $R_\mathrm{out} = 5.1 R_*$. An intermediate inclination angle of $i=60$\degr~yields $R_\mathrm{in}=1.7 R_*$ and $R_\mathrm{out} = 8.7 R_*$. These values fall within the range of \halph disc sizes surrounding Be stars, which have been observed via optical interferometry \citep{grund2006}. We note this model cannot be applied to our Epoch III and IV spectra since the assumption of optically-thin emission is not met. In these spectra the observed peak separation is controlled by radiative transfer effects related to in/outflow.

\subsection{Mass-Loss Rates}
\label{sec:mlr}

The interaction between the high-speed winds of a massive star and the local interstellar medium (ISM) can sometimes produce a bow shock that is detectable in the infrared. VES~735 produces one such bow shock feature that can be seen in the 12 $\micron$ observations from \textit{WISE} and in the 70 $\micron$ images from the \textit{Herschel Space Observatory} (\textit{HSO}). For the methods described below we assume a distance to the star of 1.45~kpc, and a spectral type of \app O8.5~V with characteristic parameters from \citet{Martins05}; effective temperature (T\textsubscript{eff}) of 32522~K, M\textsubscript{V} = -4.19, stellar radius R = 8.11 R\textsubscript{$\odot$}, wind terminal velocity ($v_{\infty}$) of 2300\kms, and a space velocity ($v_{*}$) of 10\kms. The wind terminal velocity was estimated from the extent of the P~Cygni profile of the \halph line described in Section~\ref{sec:EpochIII}, where the blue absorption of the profile reaches back to the continuum level (by eye estimate within 5 \r{A} or $\pm$115\kms). The space velocity was estimated assuming that most of the motion of the ISM surrounding VES~735 is due to the motion of gas from the blister \ion{H}{ii} region KR~140 moving at the ionized gas sound speed.

\begin{figure}
    \centering
    \hspace*{-0.0cm}  
    \vspace{0.5cm}
        \includegraphics[scale=0.4]{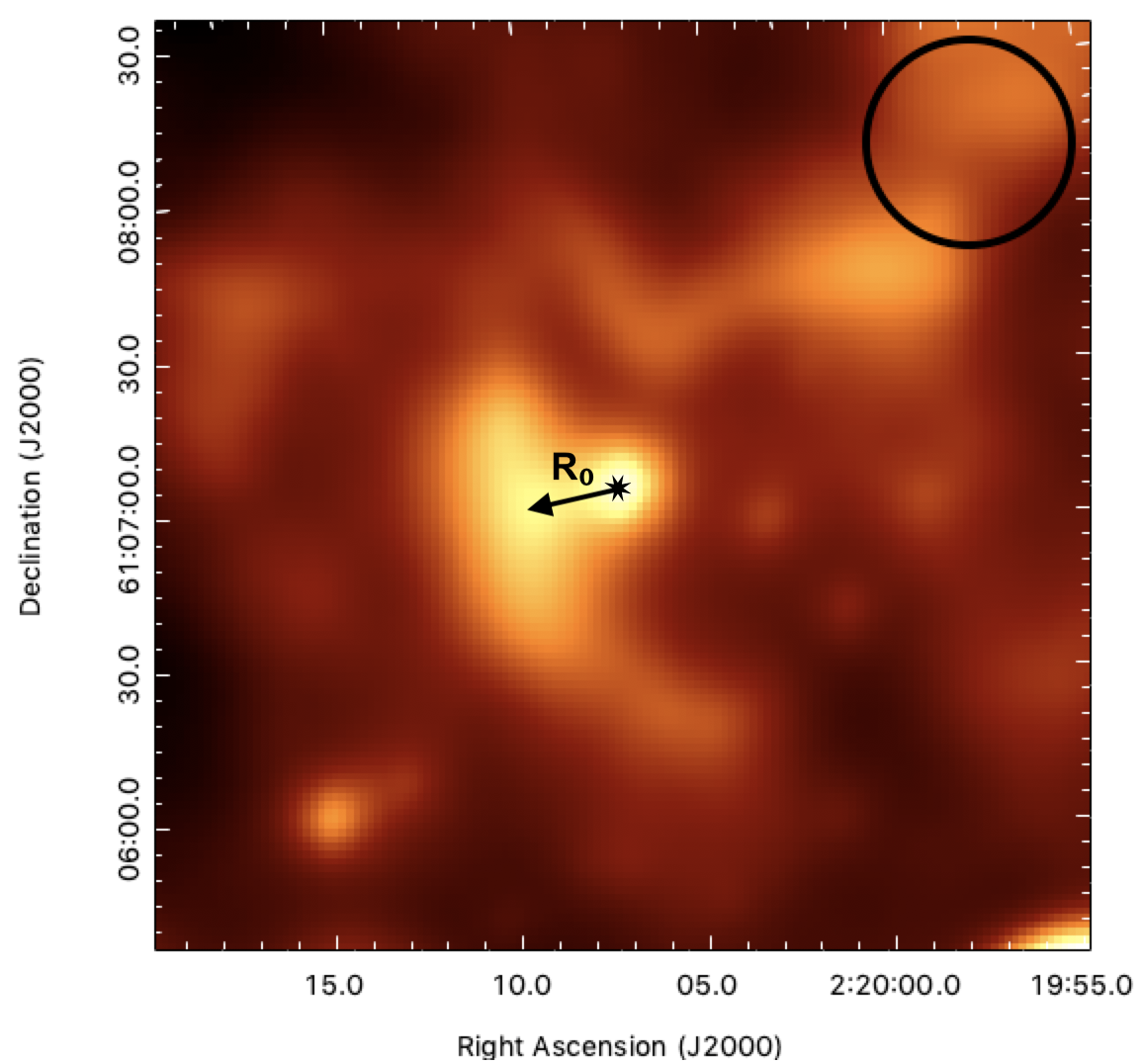}
    \includegraphics[scale=0.4]{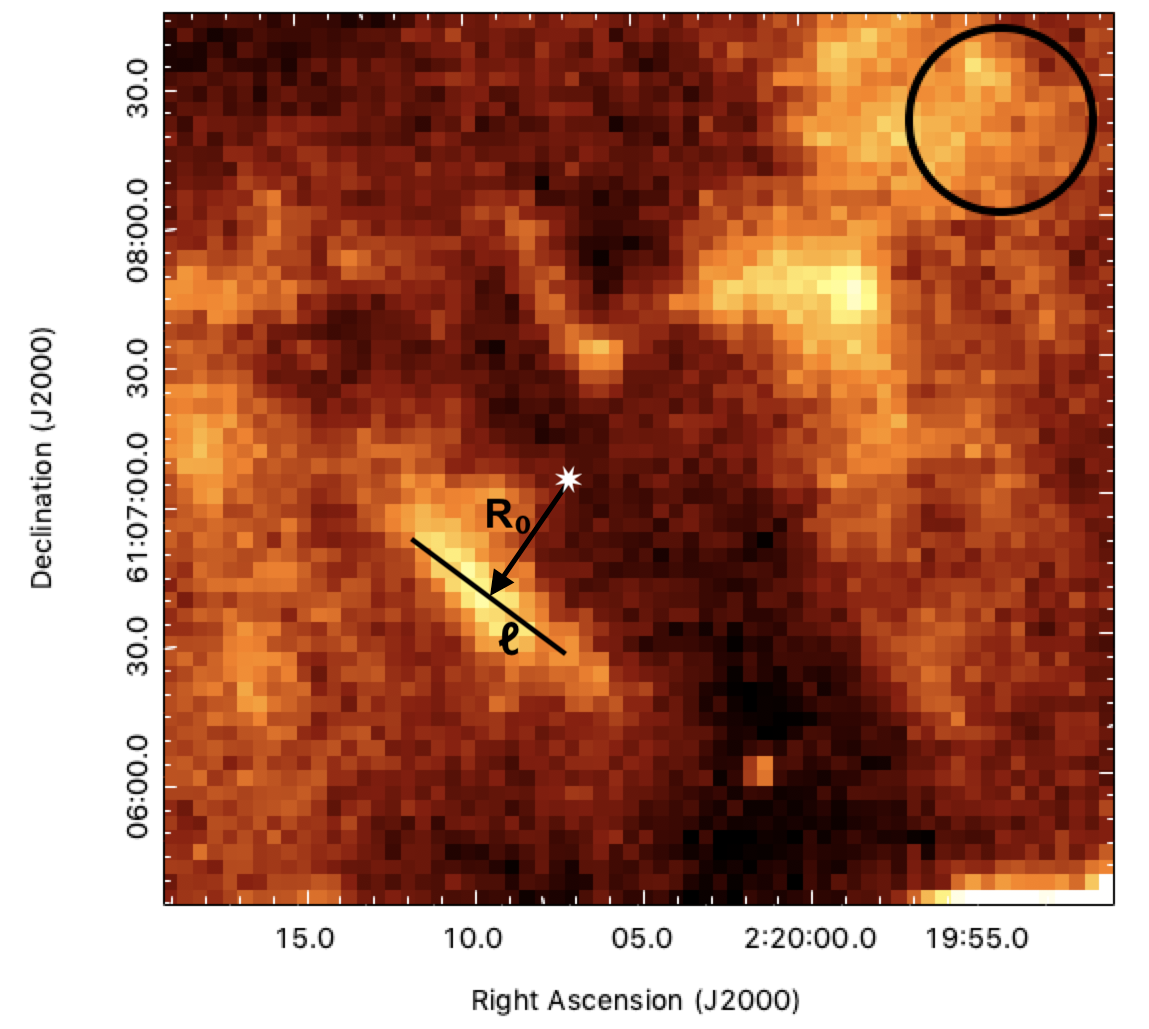}
    \caption{Top panel: \textit{WISE} 12$\micron$ image of VES~735. Black arrow is representative of the standoff distance, R\textsubscript{0}. Bottom Panel: \textit{HSO} 70$\micron$ image of VES~735. Black arrow and line represent the standoff distance, R\textsubscript{0}, and the chord diameter, $\ell$. Black circle in each panel have 0.14 pc radius.}
    \label{fig:WISE-bow}
\end{figure}

\citet{Huthoff2002} investigate the lack of detected bow shocks around high-mass X-ray binary systems (HMXBs). To define their search for bow shocks around HMXBs, they relate the stand-off distance between bow shock and star as a function of an estimated mass-loss rate, terminal wind velocity, stellar velocity (relative to the ISM), and the density of the ambient medium. Using the \textit{WISE}~12~$\micron$ images of VES~735 (top panel of Fig.~\ref{fig:WISE-bow}), we estimate the stand-off distance to be $20 \pm 5$ arcsec from the star to the peak intensity of the bow shock. A density of $23$ cm$^{-3}$ is provided in \citet{KBM99} giving a mass-loss rate of log~$\dot{M} = -8.01 \substack{+0.13 \\ -0.19}\ $M$_{\odot}$~yr$^{-1}$. The technique described in the paragraph to follow employs a statistical factor of 1.1 to account for geometric projection effects. Applying this factor we calculate a value of log~$\dot{M} = -7.93$~M$_{\odot}$~yr$^{-1}$.

The method from \citet{Kob2018} similarly relates the bow shock stand-off distance to mass-loss rate for high-mass stars; however, their method uses the 70$\micron$ \textit{HSO} images to estimate the stand-off distance, chord diameter of the bow shock and the peak intensity of the emission (see bottom panel of Fig. \ref{fig:WISE-bow}). We estimate the stand-off distance to the 70$\micron$ bow shock as the distance from the star to the peak intensity of the bow shock as $30.5 \pm 5$ arcsec. The chord length is measured through the peak intensity and spans the length of the feature, which we measure to be $45 \pm 10$ arcsec. Applying their mass-loss function, we find log~$\dot{M}$~ $=$~$-7.91\substack{+0.12 \\ -0.18}\ $M$_{\odot}$~yr$^{-1}$, which is in very good agreement with the previous technique.

Radiation-driven wind models provide another common method for determining mass-loss rates for high-mass stars by using the strength of the \halph emission. However, \citet{Lamers93} noted that this technique is unreliable for stars such as $\zeta$ Oph, where large variations in the \halph emission are due to gas in a circumstellar disc. Nevertheless, we follow their prescription of determining the mass-loss by calculating the \halph luminosity from \citet{nonLTE} non-LTE model atmospheres and equivalent width measurements of VES~735. Even when the EW is at its lowest on MJD 56901 ($EW = -2.81$ after correcting for an assumed underlying absorption of $-2.6$\r{A} \citep{KBM99}), we find log~$\dot{M}$~=~$-6.35$~M$_{\odot}~yr^{-1}$. This value is \app 1.5 dex greater than previous methods, and we agree that this method is not appropriate for stars with discs of infalling/outflowing material.

\subsection{Infrared Excess}
\label{sec:irexcess}

Oe and Be stars are known to have a near- and mid-infrared excess that is caused by free-free emission from ionized gas in a circumstellar disc \citep{geh1974,Riv13}. The striking change in the \halph emission observed in Epoch III is reflected in the infrared SED. In order to quantify this change we first re-calculate the SED slope for VES~735 during Epoch I using  \textit{Spitzer} IRAC and \textit{WISE} photometry at 3.4 and 12 $\mu$m respectively (\textit{WISE} 22~$\mu$m photometry is contaminated by emission from the nearby bow shock discussed in Section~\ref{sec:mlr}). The spectral index is $\alpha=-2.17$. For Epoch III we use \textit{NEOWISE} W1 (3.6~$\mu$m) and \textit{SOFIA} 11.2~$\micron$ photometry as proxies for IRAC 3.4~$\mu$m and \textit{WISE} 12.0~$\mu$m respectively. The \textit{NEOWISE} W1 magnitude was determined by averaging all W1 observations during the MJD 59444 data release, which are reasonably close in time to the \textit{SOFIA} observations taken on MJD 59395. The SED slope is much shallower (i.e., there is a larger infrared excess) with $\alpha = -1.29$.

\begin{figure}
   \includegraphics[scale=0.9]{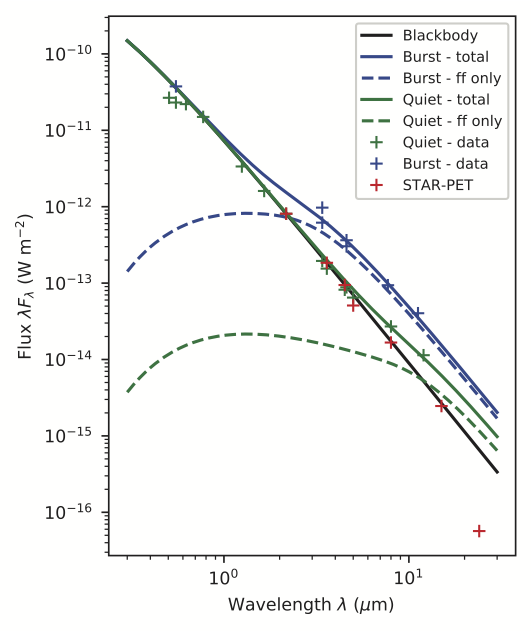}
    \caption{IR Excess Model for VES~735. Green and blue crosses indicate observations obtained during quiet (Epoch I) and burst epochs (Epoch III) respectively. The black line shows emission from a 32.5~kK blackbody, which has been scaled to match the observed emission in the $K_s$-band. The blackbody approximation agrees out to 15 $\mu$m with STAR-PET model values (red crosses) for an O8~V star (the closest available model type to O8.5~V).  The observed IR excess is modeled by adding a free-free emission spectrum associated with a hot (12~kK) ionized disc (dashed lines) to the blackbody spectrum. The free-free emission spectrum becomes optically thick at 4~$\mu$m in the burst epoch model and at 15~$\mu$m in the quiet epoch model. Solid green and blue lines show the combined emission from the star and disc. All observations have been dereddened using $A_V = 5.5$ and the extinction curve of \citet{indeb}.}
    \label{fig:SED}
\end{figure}

A simple model for optical-infrared SED of VES~735 in its quiet (Epoch I) and burst (Epoch III) epochs is shown in Fig.~\ref{fig:SED}. Quiet epoch photometry is archival 2MASS, IRAC, and \textit{WISE} data in the infrared along with \textit{Gaia} DR2 \citep{dr2} and DAO $V$-band photometry in the optical. Infrared burst epoch photometry consists of the \textit{SOFIA} observations at 7.7 and 11.2~$\mu$m, and average \textit{NEOWISE} W1 and W2 photometry from two data releases (59240 and 59444 MJD) bracketing the \textit{SOFIA} observations. The $V$-band point is the average Epoch III ZTF value. The total emission in each case is modelled as a combination of a 32.5~kK blackbody and free-free emission from a hot (12~kK) ionised disk. Each free-free spectrum could be arbitrarily scaled in intensity, and the shape of the free-free spectrum was set at short wavelengths by the temperature, and at long wavelengths by the wavelength at which the emission becomes optically thick ($\lambda_o$, $\tau(\lambda_o)=1$). We find that $\lambda_o \sim 15$~$\mu$m is appropriate in the quiet epoch and $\lambda_o \sim 4$~$\mu$m is appropriate in the burst epoch. Given the simplicity of the model, no attempt was made to formally optimise these values. The value of $\lambda_o$ is inversely proportional to the square root of the emission measure ($EM\sim n_e^2l$), where $n_e$ is the electron density, $l$ is the size/depth of the emitting region, and the electron and proton densities are assumed to be equal. The change in $\lambda_o$ from the quiet to the burst epoch, implies that $n_e$ has increased by a factor of 3.75, assuming the extent of the emitting region has not changed.

In reality the quiet to burst transition likely involves a combination of disc growth along with gas input to the disc. We can explore this a bit further using the \citet{geh1974} model in which the disc is modelled as a thin disc (or ‘shell’) extending from the star out to a radius $R_s$. Equations 2 and 3 in \citet{geh1974} present expressions for $R_s$ and for $n_e$ that include geometric factors related to the inclination angle and the degree of flattening of the disc. If we are only interested in how much $R_s$ and $n_e$ change from the quiet to burst epochs these factors cancel out and we have
\begin{equation}
\label{eqn:r_ratio}
\frac{R_{s,b}}{R_{s,q}} = \left(\frac{F_b}{F_q}\right)^{\frac{1}{2}},
\end{equation}
and
\begin{equation}
\label{eqn:ne_ratio}
\frac{n_{e,b}}{n_{e,q}} = \frac{\lambda_{o,q}}{\lambda_{o,b}}\left(\frac{F_q}{F_b} \right)^{\frac{1}{4}},
\end{equation}
where $F$ is the ratio of the extrapolated optically thick flux to the stellar flux at $\lambda_o$, and the $b$ and $q$ subscripts indicated the burst and quiet epochs respectively. For the model shown in Fig.~\ref{fig:SED} we find $F_b=5.53$ and $F_q=2.24$. Using equations (2) and (3), this means the disc size changes by a factor of 1.6 and the density by a factor of 3.0. If we increase $\lambda_o$ in the quiet epoch to 20~$\mu$m the fit to the photometric data is essentially the same and we find $F_q=3.97$. In this case the change in $R_s$ and $n_e$ is a factor of 1.2 and 4.6 respectively. We conclude that the infrared photometry supports a picture where the quiet to burst transition is dominated by a change in the disc density rather than disc size.

\section{Conclusions} 
\label{sec:conclude}

\textbf{Epoch I: }VES~735 showed a moderate infrared excess and double-peaked \halph profiles in observations spanning nearly 20 years. These observations were previously explained as being associated with a remnant accretion disc from the star's formation, and this led to the idea that the star could be classified as a new type of young stellar object, a `Herbig Oe'--type star \citep{MK18}. Based on our analysis of more recent observations, we now find that the most plausible explanation of those observations was that we were viewing a remnant \textit{decretion} disc.

\textbf{Epoch II: }The sharp increase in brightness at both visible and infrared magnitudes in mid-2015 show the first indication of a decretion event. As described by \citet{ghoreyshi2021}, the \halph EW and E/C lag the quick increase in brightness, not reaching their largest observed values until late 2021. The outer mass-reservoir takes time to grow and the overall continuum levels are enhanced during the outburst, suppressing the EW and E/C measurements until the continuum level begins to decrease.

\textbf{Epoch III: }The decline in visible brightness suggests that mass-injection into the circumstellar environment has ceased and the inner regions of the disc begin to flow back towards the star at some point between 2018 and 2019. The outer portions of the disc are still being fed with material until \app 2020, where the infrared magnitudes begin to plateau. Spectroscopic observations agree with this interpretation. The \halph profiles resemble those seen in the Oe star AzV 493 with blueshifted emission and redshifted absorption, which are commonly interpreted as arising due to infalling material. In addition the \halph EW begins to decrease with time at the end of this epoch.

\textbf{Epoch IV: }Visible magnitudes taken in early 2022 show a small spike, with infrared magnitudes slowing in their decline and potentially beginning to plateau again. The first \halph profile of the epoch had decreased in EW, suggesting that the amount of circumstellar material was truly decreasing during Epoch III. The final observations of the \halph profile are suggestive of outflow, which would also explain the increased brightness in the visible, the stabilization of the infrared magnitudes, and the increase in EW.

Conceptually VES~735 would appear to have similarities to the two-component structure posited for the B supergiant star R126 (B0.5 Ia+ [e]) by \citet{zick1985} (see their figure 7). There is a rapidly rotating star, which feeds material sporadically to a circumstellar disc, along with a strong, more spherically symmetric wind, which leads to the observed P Cygni profile, the red emission shelf, and ultimately the larger-scale bow shock structure. 

Spectroscopic and photometric observations of VES~735 have been taken periodically over the last 27 years. Observations from 1996 to 2014 showed roughly symmetric, double-peaked \halph profiles with only slight variability in line strength. Recent observations of the \halph and \he line, coupled with the visible and infrared photometry match very well with predictions from viscous decretion disc models, which have been developed to describe similar spectral and photometric variations seen in Be stars. Further monitoring of VES~735 will provide an opportunity to study disc formation and dissipation scenarios associated with the high-mass end of the OBe phenomenon.

\section*{Acknowledgements}
This research has made use of the NASA/IPAC IRSA, which is operated by the Jet Propulsion Laboratory, California Institute of Technology, under contract with the National Aeronautics and Space Administration. This research also makes use of data products from the Near-Earth Object Wide-field Infrared Survey Explorer (NEOWISE), which is a joint project of the Jet Propulsion Laboratory/California Institute of Technology and the University of Arizona. NEOWISE is funded by the National Aeronautics and Space Administration.

This work is based in part on observations made with the NASA/DLR Stratospheric Observatory for Infrared Astronomy (SOFIA). SOFIA is jointly operated by the Universities Space Research Association, Inc. (USRA), under NASA contract NNA17BF53C, and the Deutsches SOFIA Institute (DSI) under DLR contract 50 OK 0901 to the University of Stuttgart. Financial support for this work was provided by NASA through award \#08-0245 issued by USRA.

We acknowledge with thanks the variable star observations from the AAVSO International Database contributed by observers worldwide and used in this research and in particular A. Popowicz and the team at Silesian University of Technology Observatories.
We thank the following for acquisition and subsequent data reduction of VES~735 spectra: H. Kobulnicky, M.J. Lindman, and E. Cook at WIRO and APO, and D. Bohlender at DAO. We thank the anonymous referee for their careful reading, comments, and questions that greatly improved the manuscript.


\section*{Data Availability}

The majority of the data underlying this article can be accessed from the CADC (\url{https://www.cadc-ccda.hia-iha.nrc-cnrc.gc.ca/}), IRSA (\url{http:irsa.ipac.caltech.edu/}), AAVSO (\url{https://www.aavso.org/}), NEOWISE-R Single-exposure Source Table (\url{ https://www.ipac.caltech.edu/doi/irsa/10.26131/IRSA144}), and SEIP (\url{ https://www.ipac.caltech.edu/doi/irsa/10.26131/IRSA433}). The derived data generated in this research will be shared on reasonable request to the corresponding author. The \halph spectra underlying this article not publicly available will also be shared on reasonable request to the corresponding author.


\bibliographystyle{mnras}
\bibliography{VES_735} 

\begin{thebibliography}{}
\makeatletter
\relax
\def\mn@urlcharsother{\let\do\@makeother \do\$\do\&\do\#\do\^\do\_\do\%\do\~}
\def\mn@doi{\begingroup\mn@urlcharsother \@ifnextchar [ {\mn@doi@} {\mn@doi@[]}}
\def\mn@doi@[#1]#2{\def\@tempa{#1}\ifx\@tempa\@empty \href {http://dx.doi.org/#2} {doi:#2}\else \href {http://dx.doi.org/#2} {#1}\fi \endgroup}
\def\mn@eprint#1#2{\mn@eprint@#1:#2::\@nil}
\def\mn@eprint@arXiv#1{\href {http://arxiv.org/abs/#1} {{\tt arXiv:#1}}}
\def\mn@eprint@dblp#1{\href {http://dblp.uni-trier.de/rec/bibtex/#1.xml} {dblp:#1}}
\def\mn@eprint@#1:#2:#3:#4\@nil{\def\@tempa {#1}\def\@tempb {#2}\def\@tempc {#3}\ifx \@tempc \@empty \let \@tempc \@tempb \let \@tempb \@tempa \fi \ifx \@tempb \@empty \def\@tempb {arXiv}\fi \@ifundefined {mn@eprint@\@tempb}{\@tempb:\@tempc}{\expandafter \expandafter \csname mn@eprint@\@tempb\endcsname \expandafter{\@tempc}}}

\bibitem[\protect\citeauthoryear{{Auer} \& {van Blerkom}}{{Auer} \& {van Blerkom}}{1972}]{auer1972}
{Auer} L.~H.,  {van Blerkom} D.,  1972, \mn@doi [\apj] {10.1086/151777}, \href {https://ui.adsabs.harvard.edu/abs/1972ApJ...178..175A} {178, 175}

\bibitem[\protect\citeauthoryear{{Beals}}{{Beals}}{1953}]{Beals1953}
{Beals} C.~S.,  1953, Publications of the Dominion Astrophysical Observatory Victoria, \href {https://ui.adsabs.harvard.edu/abs/1953PDAO....9....1B} {9, 1}

\bibitem[\protect\citeauthoryear{{Bik}, {Kaper}  \& {Waters}}{{Bik} et~al.}{2006}]{bik}
{Bik} A.,  {Kaper} L.,   {Waters} L.~B.~F.~M.,  2006, \mn@doi [\aap] {10.1051/0004-6361:20042403}, \href {https://ui.adsabs.harvard.edu/abs/2006A&A...455..561B} {455, 561}

\bibitem[\protect\citeauthoryear{{Carciofi}, {Bjorkman}, {Otero}, {Okazaki}, {{\v{S}}tefl}, {Rivinius}, {Baade}  \& {Haubois}}{{Carciofi} et~al.}{2012}]{carciofi2012}
{Carciofi} A.~C.,  {Bjorkman} J.~E.,  {Otero} S.~A.,  {Okazaki} A.~T.,  {{\v{S}}tefl} S.,  {Rivinius} T.,  {Baade} D.,   {Haubois} X.,  2012, \mn@doi [\apjl] {10.1088/2041-8205/744/1/L15}, \href {https://ui.adsabs.harvard.edu/abs/2012ApJ...744L..15C} {744, L15}

\bibitem[\protect\citeauthoryear{{Clark}, {Tarasov}, {Okazaki}, {Roche}  \& {Lyuty}}{{Clark} et~al.}{2001}]{Clark2001}
{Clark} J.~S.,  {Tarasov} A.~E.,  {Okazaki} A.~T.,  {Roche} P.,   {Lyuty} V.~M.,  2001, \mn@doi [\aap] {10.1051/0004-6361:20011468}, \href {https://ui.adsabs.harvard.edu/abs/2001A&A...380..615C} {380, 615}

\bibitem[\protect\citeauthoryear{{Conti} \& {Leep}}{{Conti} \& {Leep}}{1974}]{cl74}
{Conti} P.~S.,  {Leep} E.~M.,  1974, \mn@doi [\apj] {10.1086/153135}, \href {http://adsabs.harvard.edu/abs/1974ApJ...193..113C} {193, 113}

\bibitem[\protect\citeauthoryear{{Cooke} \& {Rodgers}}{{Cooke} \& {Rodgers}}{2005}]{GNIRS}
{Cooke} A.,  {Rodgers} B.,  2005, in {Shopbell} P.,  {Britton} M.,   {Ebert} R.,  eds,  Astronomical Society of the Pacific Conference Series Vol. 347, Astronomical Data Analysis Software and Systems XIV. p.~514

\bibitem[\protect\citeauthoryear{{Coyne} \& {MacConnell}}{{Coyne} \& {MacConnell}}{1983}]{VAT}
{Coyne} G.~V.,  {MacConnell} D.~J.,  1983, Vatican Observatory Publications, \href {https://ui.adsabs.harvard.edu/abs/1983VatOP...2...73C} {1, 73}

\bibitem[\protect\citeauthoryear{{Crowther}}{{Crowther}}{2005}]{Crowther05}
{Crowther} P.~A.,  2005, in {Cesaroni} R.,  {Felli} M.,  {Churchwell} E.,   {Walmsley} M.,  eds,  Vol. 227, Massive Star Birth: A Crossroads of Astrophysics. pp 389--396 (\mn@eprint {arXiv} {astro-ph/0506324}), \mn@doi{10.1017/S1743921305004795}

\bibitem[\protect\citeauthoryear{{Ebbets}}{{Ebbets}}{1981}]{ebb81}
{Ebbets} D.,  1981, \mn@doi [\pasp] {10.1086/130787}, \href {https://ui.adsabs.harvard.edu/abs/1981PASP...93..119E} {93, 119}

\bibitem[\protect\citeauthoryear{{Fazio} et~al.,}{{Fazio} et~al.}{2004}]{faz2004}
{Fazio} G.~G.,  et~al., 2004, \mn@doi [\apjs] {10.1086/422843}, \href {https://ui.adsabs.harvard.edu/abs/2004ApJS..154...10F} {154, 10}

\bibitem[\protect\citeauthoryear{{Gaia Collaboration}}{{Gaia Collaboration}}{2020}]{Gaia}
{Gaia Collaboration} 2020, VizieR Online Data Catalog, \href {https://ui.adsabs.harvard.edu/abs/2020yCat.1350....0G} {p. I/350}

\bibitem[\protect\citeauthoryear{{Gaia Collaboration} et~al.,}{{Gaia Collaboration} et~al.}{2018}]{dr2}
{Gaia Collaboration} et~al., 2018, \mn@doi [\aap] {10.1051/0004-6361/201833051}, \href {https://ui.adsabs.harvard.edu/abs/2018A&A...616A...1G} {616, A1}

\bibitem[\protect\citeauthoryear{{Gehrz}, {Hackwell}  \& {Jones}}{{Gehrz} et~al.}{1974}]{geh1974}
{Gehrz} R.~D.,  {Hackwell} J.~A.,   {Jones} T.~W.,  1974, \mn@doi [\apj] {10.1086/153008}, \href {https://ui.adsabs.harvard.edu/abs/1974ApJ...191..675G} {191, 675}

\bibitem[\protect\citeauthoryear{{Ghoreyshi} \& {Carciofi}}{{Ghoreyshi} \& {Carciofi}}{2017}]{ghoreyshi2017}
{Ghoreyshi} M.~R.,  {Carciofi} A.~C.,  2017, in {Miroshnichenko} A.,  {Zharikov} S.,  {Kor{\v{c}}{\'a}kov{\'a}} D.,   {Wolf} M.,  eds,  Astronomical Society of the Pacific Conference Series Vol. 508, The B[e] Phenomenon: Forty Years of Studies. p.~323 (\mn@eprint {arXiv} {1702.06982}), \mn@doi{10.48550/arXiv.1702.06982}

\bibitem[\protect\citeauthoryear{{Ghoreyshi}, {Carciofi}, {Jones}, {Faes}, {Baade}  \& {Rivinius}}{{Ghoreyshi} et~al.}{2021}]{ghoreyshi2021}
{Ghoreyshi} M.~R.,  {Carciofi} A.~C.,  {Jones} C.~E.,  {Faes} D.~M.,  {Baade} D.,   {Rivinius} T.,  2021, \mn@doi [\apj] {10.3847/1538-4357/abdd1e}, \href {https://ui.adsabs.harvard.edu/abs/2021ApJ...909..149G} {909, 149}

\bibitem[\protect\citeauthoryear{{Golden-Marx}, {Oey}, {Lamb}, {Graus}  \& {White}}{{Golden-Marx} et~al.}{2016}]{Gold16}
{Golden-Marx} J.~B.,  {Oey} M.~S.,  {Lamb} J.~B.,  {Graus} A.~S.,   {White} A.~S.,  2016, \mn@doi [\apj] {10.3847/0004-637X/819/1/55}, \href {https://ui.adsabs.harvard.edu/abs/2016ApJ...819...55G} {819, 55}

\bibitem[\protect\citeauthoryear{{Groppi} \& {Hanson}}{{Groppi} \& {Hanson}}{1996}]{RedClass}
{Groppi} C.~E.,  {Hanson} M.~M.,  1996, \mn@doi [\pasp] {10.1086/133767}, \href {https://ui.adsabs.harvard.edu/abs/1996PASP..108..575G} {108, 575}

\bibitem[\protect\citeauthoryear{{Grundstrom} \& {Gies}}{{Grundstrom} \& {Gies}}{2006}]{grund2006}
{Grundstrom} E.~D.,  {Gies} D.~R.,  2006, \mn@doi [\apjl] {10.1086/509635}, \href {https://ui.adsabs.harvard.edu/abs/2006ApJ...651L..53G} {651, L53}

\bibitem[\protect\citeauthoryear{{Hartmann}, {Herczeg}  \& {Calvet}}{{Hartmann} et~al.}{2016}]{hartmann2016}
{Hartmann} L.,  {Herczeg} G.,   {Calvet} N.,  2016, \mn@doi [\araa] {10.1146/annurev-astro-081915-023347}, \href {https://ui.adsabs.harvard.edu/abs/2016ARA&A..54..135H} {54, 135}

\bibitem[\protect\citeauthoryear{{Herter} et~al.,}{{Herter} et~al.}{2012}]{herter}
{Herter} T.~L.,  et~al., 2012, \mn@doi [\apjl] {10.1088/2041-8205/749/2/L18}, \href {https://ui.adsabs.harvard.edu/abs/2012ApJ...749L..18H} {749, L18}

\bibitem[\protect\citeauthoryear{{Huang}}{{Huang}}{1972}]{huang1972}
{Huang} S.-S.,  1972, \mn@doi [\apj] {10.1086/151309}, \href {https://ui.adsabs.harvard.edu/abs/1972ApJ...171..549H} {171, 549}

\bibitem[\protect\citeauthoryear{{Huthoff} \& {Kaper}}{{Huthoff} \& {Kaper}}{2002}]{Huthoff2002}
{Huthoff} F.,  {Kaper} L.,  2002, \mn@doi [\aap] {10.1051/0004-6361:20011793}, \href {https://ui.adsabs.harvard.edu/abs/2002A&A...383..999H} {383, 999}

\bibitem[\protect\citeauthoryear{{Indebetouw} et~al.,}{{Indebetouw} et~al.}{2005}]{indeb}
{Indebetouw} R.,  et~al., 2005, \mn@doi [\apj] {10.1086/426679}, \href {https://ui.adsabs.harvard.edu/abs/2005ApJ...619..931I} {619, 931}

\bibitem[\protect\citeauthoryear{{Kerton}, {Ballantyne}  \& {Martin}}{{Kerton} et~al.}{1999}]{KBM99}
{Kerton} C.~R.,  {Ballantyne} D.~R.,   {Martin} P.~G.,  1999, \mn@doi [\aj] {10.1086/300858}, \href {https://ui.adsabs.harvard.edu/abs/1999AJ....117.2485K} {117, 2485}

\bibitem[\protect\citeauthoryear{{Kobulnicky}, {Chick}  \& {Povich}}{{Kobulnicky} et~al.}{2019}]{Kob2018}
{Kobulnicky} H.~A.,  {Chick} W.~T.,   {Povich} M.~S.,  2019, \mn@doi [\aj] {10.3847/1538-3881/ab2716}, \href {https://ui.adsabs.harvard.edu/abs/2019AJ....158...73K} {158, 73}

\bibitem[\protect\citeauthoryear{{Kochanek} et~al.,}{{Kochanek} et~al.}{2017}]{Kochanek}
{Kochanek} C.~S.,  et~al., 2017, \mn@doi [\pasp] {10.1088/1538-3873/aa80d9}, \href {https://ui.adsabs.harvard.edu/abs/2017PASP..129j4502K} {129, 104502}

\bibitem[\protect\citeauthoryear{{Kraus} et~al.,}{{Kraus} et~al.}{2008}]{kraus2008}
{Kraus} S.,  et~al., 2008, \mn@doi [\aap] {10.1051/0004-6361:200809946}, \href {https://ui.adsabs.harvard.edu/abs/2008A&A...489.1157K} {489, 1157}

\bibitem[\protect\citeauthoryear{{Kurucz}}{{Kurucz}}{1993}]{kurucz93}
{Kurucz} R.~L.,  1993, VizieR Online Data Catalog, \href {https://ui.adsabs.harvard.edu/abs/1993yCat.6039....0K} {p. VI/39}

\bibitem[\protect\citeauthoryear{{Lada}}{{Lada}}{1987}]{lada87}
{Lada} C.~J.,  1987, in {Peimbert} M.,  {Jugaku} J.,  eds,  Vol. 115, Star Forming Regions. p.~1

\bibitem[\protect\citeauthoryear{{Laher}, {Gorjian}, {Rebull}, {Masci}, {Fowler}, {Helou}, {Kulkarni}  \& {Law}}{{Laher} et~al.}{2012}]{APT}
{Laher} R.~R.,  {Gorjian} V.,  {Rebull} L.~M.,  {Masci} F.~J.,  {Fowler} J.~W.,  {Helou} G.,  {Kulkarni} S.~R.,   {Law} N.~M.,  2012, \mn@doi [\pasp] {10.1086/666883}, \href {https://ui.adsabs.harvard.edu/abs/2012PASP..124..737L} {124, 737}

\bibitem[\protect\citeauthoryear{{Lamers} \& {Leitherer}}{{Lamers} \& {Leitherer}}{1993}]{Lamers93}
{Lamers} H. J.~G.~L.~M.,  {Leitherer} C.,  1993, \mn@doi [\apj] {10.1086/172960}, \href {https://ui.adsabs.harvard.edu/abs/1993ApJ...412..771L} {412, 771}

\bibitem[\protect\citeauthoryear{{Lanz} \& {Hubeny}}{{Lanz} \& {Hubeny}}{2003}]{nonLTE}
{Lanz} T.,  {Hubeny} I.,  2003, \mn@doi [\apjs] {10.1086/374373}, \href {https://ui.adsabs.harvard.edu/abs/2003ApJS..146..417L} {146, 417}

\bibitem[\protect\citeauthoryear{{Lejeune}, {Cuisinier}  \& {Buser}}{{Lejeune} et~al.}{1997}]{lejeune97}
{Lejeune} T.,  {Cuisinier} F.,   {Buser} R.,  1997, \mn@doi [\aaps] {10.1051/aas:1997373}, \href {https://ui.adsabs.harvard.edu/abs/1997A&AS..125..229L} {125, 229}

\bibitem[\protect\citeauthoryear{{Mainzer} et~al.,}{{Mainzer} et~al.}{2011}]{NeoWise}
{Mainzer} A.,  et~al., 2011, \mn@doi [\apj] {10.1088/0004-637X/731/1/53}, \href {https://ui.adsabs.harvard.edu/abs/2011ApJ...731...53M} {731, 53}

\bibitem[\protect\citeauthoryear{{Mainzer} et~al.,}{{Mainzer} et~al.}{2014}]{NEOWISE14}
{Mainzer} A.,  et~al., 2014, \mn@doi [\apj] {10.1088/0004-637X/792/1/30}, \href {https://ui.adsabs.harvard.edu/abs/2014ApJ...792...30M} {792, 30}

\bibitem[\protect\citeauthoryear{{Ma{\'\i}z Apell{\'a}niz} et~al.,}{{Ma{\'\i}z Apell{\'a}niz} et~al.}{2016}]{als}
{Ma{\'\i}z Apell{\'a}niz} J.,  et~al., 2016, \mn@doi [\apjs] {10.3847/0067-0049/224/1/4}, \href {https://ui.adsabs.harvard.edu/abs/2016ApJS..224....4M} {224, 4}

\bibitem[\protect\citeauthoryear{{Mamajek}}{{Mamajek}}{2023}]{mem23}
{Mamajek} E.,  2023, A Modern Mean Dwarf Stellar Color and Effective Temperature Sequence, ver. 2022.04.16, \url{https://www.pas.rochester.edu/~emamajek/EEM_dwarf_UBVIJHK_colors_Teff.txt}

\bibitem[\protect\citeauthoryear{{Marlborough}, {Zijlstra}  \& {Waters}}{{Marlborough} et~al.}{1997}]{marl97}
{Marlborough} J.~M.,  {Zijlstra} J.~W.,   {Waters} L.~B.~F.~M.,  1997, \aap, \href {https://ui.adsabs.harvard.edu/abs/1997A&A...321..867M} {321, 867}

\bibitem[\protect\citeauthoryear{{Marshall} \& {Kerton}}{{Marshall} \& {Kerton}}{2018}]{MK18}
{Marshall} B.,  {Kerton} C.~R.,  2018, \mn@doi [Research Notes of the American Astronomical Society] {10.3847/2515-5172/aaf505}, \href {https://ui.adsabs.harvard.edu/abs/2018RNAAS...2..221M} {2, 221}

\bibitem[\protect\citeauthoryear{{Martins}, {Schaerer}  \& {Hillier}}{{Martins} et~al.}{2005}]{Martins05}
{Martins} F.,  {Schaerer} D.,   {Hillier} D.~J.,  2005, \mn@doi [\aap] {10.1051/0004-6361:20042386}, \href {https://ui.adsabs.harvard.edu/abs/2005A&A...436.1049M} {436, 1049}

\bibitem[\protect\citeauthoryear{{Masci} et~al.,}{{Masci} et~al.}{2019}]{Masci}
{Masci} F.~J.,  et~al., 2019, \mn@doi [\pasp] {10.1088/1538-3873/aae8ac}, \href {https://ui.adsabs.harvard.edu/abs/2019PASP..131a8003M} {131, 018003}

\bibitem[\protect\citeauthoryear{{Massey}}{{Massey}}{1997}]{Massey}
{Massey} P.,  1997, {Instruction Manual for IRAF Software Package}, \url{http://www.iac.es/sieinvens/SINFIN/CursoIraf/Cap_references.php}

\bibitem[\protect\citeauthoryear{{McLaughlin}}{{McLaughlin}}{1961}]{McL61}
{McLaughlin} D.~B.,  1961, \jrasc, \href {https://ui.adsabs.harvard.edu/abs/1961JRASC..55...73M} {55, 73}

\bibitem[\protect\citeauthoryear{{Negueruela}, {Steele}  \& {Bernabeu}}{{Negueruela} et~al.}{2004}]{neg2004}
{Negueruela} I.,  {Steele} I.~A.,   {Bernabeu} G.,  2004, \mn@doi [Astronomische Nachrichten] {10.1002/asna.200310258}, \href {https://ui.adsabs.harvard.edu/abs/2004AN....325..749N} {325, 749}

\bibitem[\protect\citeauthoryear{{Niemel{\"a}} \& {M{\'e}ndez}}{{Niemel{\"a}} \& {M{\'e}ndez}}{1974}]{niemela}
{Niemel{\"a}} V.~S.,  {M{\'e}ndez} R.~H.,  1974, \mn@doi [\apjl] {10.1086/181383}, \href {https://ui.adsabs.harvard.edu/abs/1974ApJ...187L..23N} {187, L23}

\bibitem[\protect\citeauthoryear{{Oey} et~al.,}{{Oey} et~al.}{2023}]{oey2023}
{Oey} M.~S.,  et~al., 2023, \mn@doi [\apj] {10.3847/1538-4357/acb690}, \href {https://ui.adsabs.harvard.edu/abs/2023ApJ...947...27O} {947, 27}

\bibitem[\protect\citeauthoryear{{Okazaki}}{{Okazaki}}{1991}]{1991PASJ...43...75O}
{Okazaki} A.~T.,  1991, \pasj, \href {https://ui.adsabs.harvard.edu/abs/1991PASJ...43...75O} {43, 75}

\bibitem[\protect\citeauthoryear{{Okazaki}}{{Okazaki}}{1997}]{Okazaki97}
{Okazaki} A.~T.,  1997, \aap, \href {https://ui.adsabs.harvard.edu/abs/1997A&A...318..548O} {318, 548}

\bibitem[\protect\citeauthoryear{{Panagia}}{{Panagia}}{1973}]{Panagia73}
{Panagia} N.,  1973, \mn@doi [\aj] {10.1086/111498}, \href {https://ui.adsabs.harvard.edu/abs/1973AJ.....78..929P} {78, 929}

\bibitem[\protect\citeauthoryear{{Paul}, {Shruthi}  \& {Subramaniam}}{{Paul} et~al.}{2017}]{Paul17}
{Paul} K.~T.,  {Shruthi} S.~B.,   {Subramaniam} A.,  2017, \mn@doi [Journal of Astrophysics and Astronomy] {10.1007/s12036-017-9426-0}, \href {https://ui.adsabs.harvard.edu/abs/2017JApA...38....6P} {38, 6}

\bibitem[\protect\citeauthoryear{{R{\'\i}mulo} et~al.,}{{R{\'\i}mulo} et~al.}{2018}]{rimulo2018}
{R{\'\i}mulo} L.~R.,  et~al., 2018, \mn@doi [\mnras] {10.1093/mnras/sty431}, \href {https://ui.adsabs.harvard.edu/abs/2018MNRAS.476.3555R} {476, 3555}

\bibitem[\protect\citeauthoryear{{Rivinius}, {Carciofi}  \& {Martayan}}{{Rivinius} et~al.}{2013}]{Riv13}
{Rivinius} T.,  {Carciofi} A.~C.,   {Martayan} C.,  2013, \mn@doi [\aapr] {10.1007/s00159-013-0069-0}, \href {https://ui.adsabs.harvard.edu/abs/2013A&ARv..21...69R} {21, 69}

\bibitem[\protect\citeauthoryear{{Robitaille}, {Whitney}, {Indebetouw}  \& {Wood}}{{Robitaille} et~al.}{2007}]{rob07}
{Robitaille} T.~P.,  {Whitney} B.~A.,  {Indebetouw} R.,   {Wood} K.,  2007, \mn@doi [\apjs] {10.1086/512039}, \href {https://ui.adsabs.harvard.edu/\#abs/2007ApJS..169..328R} {169, 328}

\bibitem[\protect\citeauthoryear{{Shappee} et~al.,}{{Shappee} et~al.}{2014}]{Shappee}
{Shappee} B.~J.,  et~al., 2014, \mn@doi [\apj] {10.1088/0004-637X/788/1/48}, \href {https://ui.adsabs.harvard.edu/abs/2014ApJ...788...48S} {788, 48}

\bibitem[\protect\citeauthoryear{{Skrutskie} et~al.,}{{Skrutskie} et~al.}{2006}]{skr06}
{Skrutskie} M.~F.,  et~al., 2006, \mn@doi [\aj] {10.1086/498708}, \href {https://ui.adsabs.harvard.edu/abs/2006AJ....131.1163S} {131, 1163}

\bibitem[\protect\citeauthoryear{{Sota}, {Ma{\'\i}z Apell{\'a}niz}, {Walborn}, {Alfaro}, {Barb{\'a}}, {Morrell}, {Gamen}  \& {Arias}}{{Sota} et~al.}{2011}]{Sota11}
{Sota} A.,  {Ma{\'\i}z Apell{\'a}niz} J.,  {Walborn} N.~R.,  {Alfaro} E.~J.,  {Barb{\'a}} R.~H.,  {Morrell} N.~I.,  {Gamen} R.~C.,   {Arias} J.~I.,  2011, \mn@doi [\apjs] {10.1088/0067-0049/193/2/24}, \href {https://ui.adsabs.harvard.edu/abs/2011ApJS..193...24S} {193, 24}

\bibitem[\protect\citeauthoryear{{Sota}, {Ma{\'\i}z Apell{\'a}niz}, {Morrell}, {Barb{\'a}}, {Walborn}, {Gamen}, {Arias}  \& {Alfaro}}{{Sota} et~al.}{2014}]{Sota14}
{Sota} A.,  {Ma{\'\i}z Apell{\'a}niz} J.,  {Morrell} N.~I.,  {Barb{\'a}} R.~H.,  {Walborn} N.~R.,  {Gamen} R.~C.,  {Arias} J.~I.,   {Alfaro} E.~J.,  2014, \mn@doi [\apjs] {10.1088/0067-0049/211/1/10}, \href {https://ui.adsabs.harvard.edu/abs/2014ApJS..211...10S} {211, 10}

\bibitem[\protect\citeauthoryear{{Thompson}}{{Thompson}}{1984}]{Thompson84}
{Thompson} R.~I.,  1984, \mn@doi [\apj] {10.1086/162287}, \href {https://ui.adsabs.harvard.edu/abs/1984ApJ...283..165T} {283, 165}

\bibitem[\protect\citeauthoryear{{Watson} et~al.,}{{Watson} et~al.}{2008}]{Watson08}
{Watson} C.,  et~al., 2008, \mn@doi [\apj] {10.1086/588005}, \href {https://ui.adsabs.harvard.edu/abs/2008ApJ...681.1341W} {681, 1341}

\bibitem[\protect\citeauthoryear{{Wright} et~al.,}{{Wright} et~al.}{2010}]{Wright}
{Wright} E.~L.,  et~al., 2010, \mn@doi [\aj] {10.1088/0004-6256/140/6/1868}, \href {https://ui.adsabs.harvard.edu/abs/2010AJ....140.1868W} {140, 1868}

\bibitem[\protect\citeauthoryear{{Young} et~al.,}{{Young} et~al.}{2012}]{Young}
{Young} E.~T.,  et~al., 2012, \mn@doi [\apjl] {10.1088/2041-8205/749/2/L17}, \href {https://ui.adsabs.harvard.edu/abs/2012ApJ...749L..17Y} {749, L17}

\bibitem[\protect\citeauthoryear{{Zickgraf}, {Wolf}, {Stahl}, {Leitherer}  \& {Klare}}{{Zickgraf} et~al.}{1985}]{zick1985}
{Zickgraf} F.~J.,  {Wolf} B.,  {Stahl} O.,  {Leitherer} C.,   {Klare} G.,  1985, \aap, \href {https://ui.adsabs.harvard.edu/abs/1985A&A...143..421Z} {143, 421}

\bibitem[\protect\citeauthoryear{{Zorec} \& {Briot}}{{Zorec} \& {Briot}}{1997}]{Zorec97}
{Zorec} J.,  {Briot} D.,  1997, \aap, \href {https://ui.adsabs.harvard.edu/abs/1997A&A...318..443Z} {318, 443}

\bibitem[\protect\citeauthoryear{{{\v{S}}tefl}, {Okazaki}, {Rivinius}  \& {Baade}}{{{\v{S}}tefl} et~al.}{2007}]{Stefl07}
{{\v{S}}tefl} S.,  {Okazaki} A.~T.,  {Rivinius} T.,   {Baade} D.,  2007, in {Okazaki} A.~T.,  {Owocki} S.~P.,   {Stefl} S.,  eds,  Astronomical Society of the Pacific Conference Series Vol. 361, Active OB-Stars: Laboratories for Stellare and Circumstellar Physics. p.~274

\makeatother
\end{thebibliography}



\appendix

\section{Full Program of Spectra}
Table \ref{tab:appendix} lists the spectroscopic observations used for this study including the observatory codes (detailed in Section \ref{sec:obs}), Modified Juilan Date (MJD), signal-to-noise ratio (S/N), spectral resolution (R), spectral range in \r{A}, and program ID when available. Note that the precise dates of the DDO observations are not know and have been estimated to the 15th of the month in which we know the observation occurred. 

\begin{table*}
    \centering
    \caption{Spectroscopic Observations of VES~735}
    \begin{tabular}[width = \textwidth]{c|c|c|c|c|c}
    \hline
Observatory	&	MJD Start	&	S/N	&	R	&		Wavelength Range (\r{A})		&	Program ID	\\
\hline
\hline
DDO	&	50371	   &	&	13000	&	6500	--	6700	&		\\
DDO	&	50463	   &	&	13000	&	6500	--	6700	&		\\
DAO	&	50711.46707	   &	&	9400	&	6480	--	6690	&		\\
DAO	&	50712.48395	   &	&	9400	&	6480	--	6690	&		\\
DDO	&	50797	   &	&	13000	&	6500	--	6700	&		\\
WIRO	&	56533.46771	&	80	&	3000	&	5462	--	6750	&\\
WIRO	&	56626.10578	&	90	&	3000	&	5320	--	6617	&\\
WIRO	&	56901.43825	&	70	&	3000	&	5320	--	6625	&\\
Gemini	&	57677.60123	     &	30	&	1800	&	18769	--	24782	&	GN-2016B-FT-17	\\
Gemini	&	57677.60148	&	30	&	1800	&	18769	--	24782	&	GN-2016B-FT-17	\\
Gemini	&	57677.60171	&	30	&	1800	&	18769	--	24782	&	GN-2016B-FT-17	\\
Gemini	&	57677.60197	&	30	&	1800	&	18769	--	24782	&	GN-2016B-FT-17	\\
APO	&	58786.24006	&	200 &	3100	&	5717	--	6877	&\\
DAO	&	59535.29954 &   35	&	9400	&	6490	--	6751	&	DAO182 2021D4	\\
DAO	&	59535.32043	&	35	&	9400	&	6490	--	6751	&	DAO182 2021D4	\\
DAO	&	59542.24524	&	40	&	9400	&	6490	--	6751	&	DAO182 2021D4	\\
DAO	&	59542.26615	&	40	&	9400	&	6490	--	6751	&	DAO182 2021D4	\\
DAO	&	59567.23803	&	50	&	9400	&	6489	--	6750	&	DAO182 2021D4	\\
DAO	&	59567.25894	&	50	&	9400	&	6489	--	6750	&	DAO182 2021D4	\\
DAO	&	59569.15650	&	30	&	9400	&	6489	--	6750	&	DAO182 2021D4	\\
DAO	&	59569.17740	&	40	&	9400	&	6489	--	6750	&	DAO182 2021D4	\\
DAO	&	59569.19830	&	40	&	9400	&	6489	--	6750	&	DAO182 2021D4	\\
APO	&	59728.44620	&	20	&	33000	&	6308	--	6895	&\\
DAO &	59867.35610	&	35	&	9400	&	6476	--	6737	&	DAO182 2022D4	\\
DAO	&	59867.37745	&	35	&	9400	&	6476	--	6737	&	DAO182 2022D4	\\
DAO	&	59867.39836 &	30	&	9400	&	6476	--	6737	&	DAO182 2022D4	\\
DAO	&	59867.41925	&	25	&	9400	&	6476	--	6737	&	DAO182 2022D4	\\
DAO	&	59867.44015	&	30	&	9400	&	6476	--	6737	&	DAO182 2022D4	\\
DAO	&	59867.46105	&	30	&	9400	&	6476	--	6737	&	DAO182 2022D4	\\
DAO	&	59955.13134	&	12	&	9400	&	6476	--	6737	&	DAO182 2023A3	\\
DAO	&	59955.15223	&	12	&	9400	&	6476	--	6737	&	DAO182 2023A3	\\
DAO	&	59955.17314	&	12	&	9400	&	6476	--	6737	&	DAO182 2023A3	\\
\hline
\hline
    \end{tabular}
    \label{tab:appendix}
\end{table*}

\bsp	
\label{lastpage}
\end{document}